\documentclass[twocolumn]{revtex4}
\usepackage{bm}
\usepackage{amsmath}
\usepackage{amssymb}
\usepackage{graphicx}
\usepackage{epsfig}
\usepackage{epstopdf}
\usepackage{color}

\begin{document}

\title{Shaping field correlations with quantum antennas}

\author{A. Mikhalychev$^1$, D. Mogilevtsev$^{1*}$, G. Ya. Slepyan$^2$, I. Karuseichyk$^1$, G. Buchs$^3$, D. L. Boiko$^3$, A. Boag$^2$}
\affiliation{$^1$Institute of Physics, Belarus National Academy of Sciences, Nezavisimosti Ave. 68, Minsk 220072 Belarus;\\
$^2$School of Electrical Engineering, Tel Aviv University, Tel Aviv 69978, Israel;\\
$^3$Centre Suisse d'Electronique et de Microtechnique (CSEM), Jaquet-Droz 1, 2002 Neuch\^{a}tel, Switzerland.\\
$^*$ $\mathrm{the\ corresponding\ author:}$ d.mogilevtsev@ifanbel.bas-net.by}


\begin{abstract}
Quantum antennas can shape the spatial entanglement of emitted photons originating from specific initial
non-Dicke entangled states of antenna emitters. In contrast to a classical antenna, a quantum antenna might not be affecting
the amplitudes and intensities distribution of the field, but only shaping the second and higher order correlations.  The shape and
directivity of the correlations can be optimized using quantum state inference techniques.
The character of the correlations can also be  controlled by changing both the geometry and the initial state of the antenna.
Positive and negative correlated twin-photons, as well as multi-photons entangled states can be produced from the same antenna for different initial states of the emitters. Our approach to antenna design can find applications in imaging and high-precision sensing, as well as in the development of an emitter-field interface for quantum information processing.
\end{abstract}

\date{\today}

\maketitle

\section{Introduction}

Classical antennas are devices transforming radio-waves from free-space to a guiding device and vice versa \cite{balan}. The radiative properties of such antennas are characterized by the angular distribution  of the radiated field and intensity (field and power radiation patterns). In the case of  classical radiation, the power radiation pattern formed by interference effects is equal to the squared modulus of the field radiation pattern divided by the doubled characteristic impedance of the medium  \cite{balan}.  However, this simple picture does not hold for non-classical states of emitted radiation. For example, the field  radiation pattern can vanish, whereas the power radiation pattern has a finite non-zero value. Generally,  higher-order correlation functions of the non-classical field cannot be expressed through lower-order ones.

Recent progress in nanofabrication opened a way for the design and implementation of nanoantennas operating in the terahertz, infrared, and visible spectral ranges \cite{biag,novot,alu,greffer}. In spite of the quantum origin of charge carriers transport inside the antennas, the emitted field was commonly considered as being classical. However, the use of the quantum properties of light and generalization of the concept of antennas for the quantum case \cite{boag2013,mokhl,slep2010,slep2012,slep2016} open far richer possibilities for controlling and shaping the emitted field (\textit{e.g.} directive light squeezing via antenna emission \cite{slep2016}). Note that light squeezing can be achieved not only by arranging emitters, but also by engineering the initial state of the antenna. It is well known that an entangled state of emitters can lead to entanglement of the emitted photons (\textit{i.e.} the state of the field can be mapped into the emitters state and vice versa). This effect was suggested as a basis for a quantum memory device capable to store entangled states of light \cite{gisin2011,zuk2012}. Furthermore, entanglement of emitters in antennas can lead to intensity distributions otherwise impossible to reach with the factorized initial states of the antenna emitters \cite{agarwal2011}, to sub-Rayleigh imaging and superresolution \cite{boto,rozema}, as well as to superbunching \cite{agarwal2015b}. Until now, quantum features in the field emitted by an antenna were mostly considered for some well known initial states independently of the actual antenna geometry (\textit{e.g.}, symmetric Dicke states were usually considered \cite{agarwal2011}). On the other hand, so called ``timed" Dicke states bear information about the location of emitters \cite{scully2006} and provide a special quantum mechanism that introduces a non-reciprocity of the antenna \cite{boag2013}).

Here, we introduce a method to design an initial state in a quantum antenna in order to shape the emitted field correlation functions. The non-classicality of the antenna's radiation is revealed through a measurement of the higher-order correlation functions. Note that such a measurement constitutes a convenient imaging tool \cite{cassano} that enables to reach superresolution \cite{oppel,classen,supertwin}.
The approach introduced here is similar to the one usually implemented in quantum state tomography. In the same spirit, one can  optimize the directivity of the correlation functions of the radiation produced by a quantum antenna. For example, by optimizing the second-order correlation function of the two-particle entangled state of an equispaced linear antenna array, we can produce  photon pairs that are strongly correlated in momentum. Interestingly, we find that both co-directional and contra-directional correlations are possible for the same spatial antenna design, but with different initial states. The same approach is also valid for multi-particles antenna states and higher-order correlation functions. In particular, we show that some initial states lead to a strong suppression of the radiation in the far-field zone, reproducing a classical effect of ``non-radiative source" \cite{balan,wolf}. Additionally, we show that in some cases, the quantum correlations of the antenna field can be captured  with a semiclassical model of the emitter-field interaction.

The outline of the paper is as follows. In the second Section, the antenna model is introduced. In the third Section, we describe the procedure for designing the antenna state with the required correlation functions. In the fourth Section, the long-time field state is considered  for providing guidelines for the field shaping. The fifth Section discusses the example of co- and contra-directional twin-photon propagation, and the sixth Section considers a suppression of the far-field radiation. Finally, the seventh Section discusses an application of a semiclassical approach in the description of the field emitted by a quantum antenna.

\section{Antenna model}

\begin{figure}[htb]

  \includegraphics[width=0.75\linewidth]{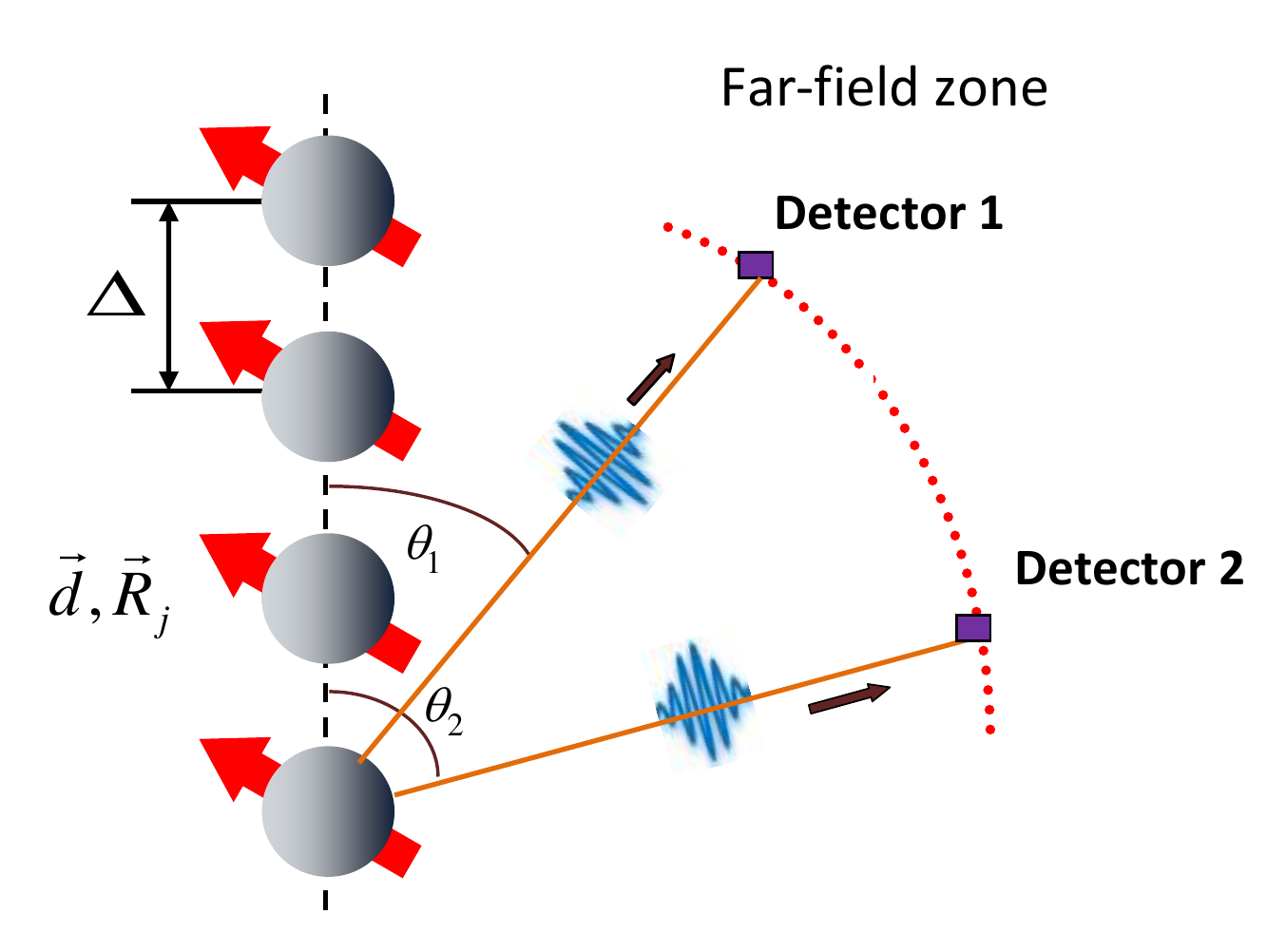}
\caption{Schematics of an equispaced linear array antenna.  The second-order
correlations can be detected by measuring simultaneous counts at detectors $D_1$ and $D_2$.  The red arrows represent the TLS dipole moments of the antenna. The dipole moments of every TLS are the same} \label{fig1}
\end{figure}

As a model system for a quantum antenna, we consider a chain of $N$ identical non-interacting two-level emitters with the same dipole moments ${\vec d}$ positioned along the same axis at points ${\vec R}_j$ (see Fig.~\ref{fig1}).  Omitting the time-dependence factor, which is common for all emitters,  the positive-frequency field operator part  that gives non-zero contribution to the normally ordered correlation functions and describes the spatial field distribution at the point ${\vec r}$ in the far field zone, reads:
\begin{eqnarray}
\vec{E}({\vec r})\propto A({\vec r})=\sum\limits_{j=1}^N\frac{{\vec n}\times[{\vec n}\times {\vec d}]}{|{\vec r}|}
\exp\{i\omega(|{\vec R}_j-{\vec r}|/c)\}\sigma^-_j,
\label{arop}
\end{eqnarray}
where $\sigma^-_j=|-_j\rangle\langle +_j|$ is the lowering operator for the $j$-th two-level system (TLS) with upper(lower) levels described by the vectors $|\pm_j\rangle$ , and ${\vec n}$ is the unit vector from an emitter to the observation point; $\omega$ is the TLS transition frequency. For what follows, we label the right-hand side in Eq. (\ref{arop}) as  the array factor operator $A({\vec r})$. Generically, the design of an antenna consists in finding the positions ${\vec R}_j$ of individual TLS elements, and in defining the initial density matrix of the antenna $\rho$ in  a way  to achieve the required values of simultaneous correlation function of the order $n$ in some sets $\{l\}$  of directions $\{{\vec r}_{k,l}\}$, $k=1\ldots n$:
\begin{equation}
G^{(n)}({\vec r}_{1,l}\ldots{\vec r}_{n,l})=\langle\left[\prod\limits_{k=1}^nA({\vec r}_{k,l})\right]^{\dagger}\prod\limits_{k=1}^nA({\vec r}_{k,l})\rangle.
\label{g}
\end{equation}
Thus, the index $l$ in Eq. (\ref{g}) labels different spatial arrangements of the $n$ detectors.   These functions can be measured by placing photon detectors in given directions and by recording the coincident counts (for example, the scheme for measuring $G^{(2)}$ is depicted  in Fig.~\ref{fig1}). Note that for a conventional classical antenna, the radiation pattern and all correlation functions are entirely defined by the average field amplitude $\langle E({\vec r})\rangle$. However, in the quantum case the situation is different. For example, for all TLSs being either in the excited or ground state, $\langle E({\vec r})\rangle=0$ for an arbitrary ${\vec r}$, whereas one can have $G^{(n)}\neq 0$.

\begin{figure}[htb]
\includegraphics[width=0.75\linewidth]{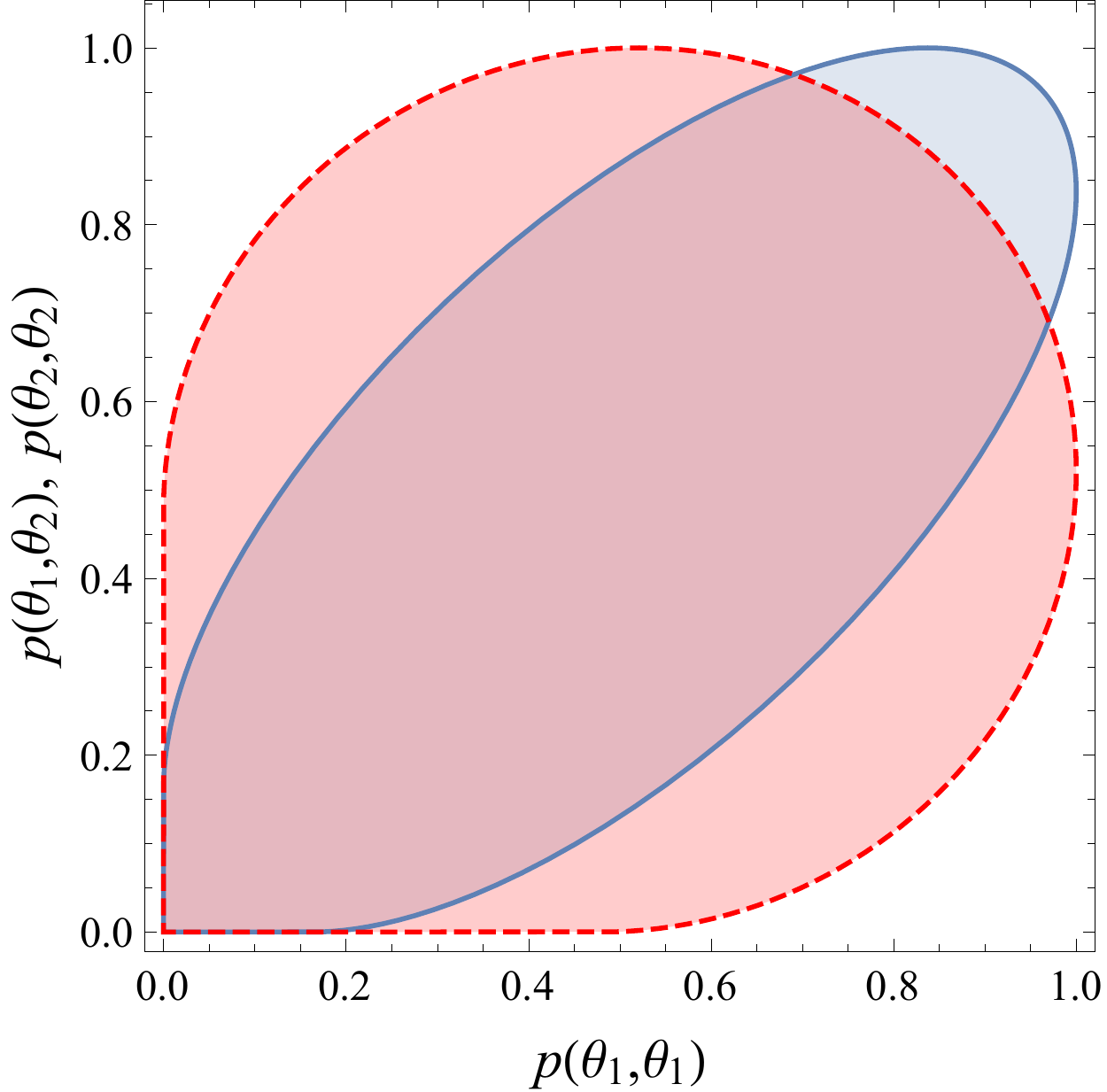}
\caption{An example of  accessible regions of target probabilities for shaping $G^{(2)}$ by a set of $N=20$ equidistant TLSs antenna depicted in Fig.~\ref{fig1} for the angles $\cos\theta_1 = 0$, $\cos\theta_2 = 0.05$ and two-excitations pure states of the form given in Eq. (\ref{psi1}). The dashed curve delimits the region of available probabilities $p(\theta_1,\theta_2)$ versus   $p(\theta_1,\theta_1)$; the solid line delimits the region of available $p(\theta_2,\theta_2)$ versus $p(\theta_1,\theta_1)$. The targeted probabilities $p_l$ are normalized by $p_l^{high}$.} \label{fig1b}
\end{figure}

\section{State estimation for antennas}

The problem of antenna state design can be formulated as a state estimation problem in the following way. We specify a finite number of discrete sets of spatial observation points for which we will perform the antenna design
$\{\vec r_{k,l}\}$, and re-write Eq.(\ref{g}) as:
\begin{equation}
p_l=Tr\{\Pi_{l}\rho\}, \quad
\Pi_{l}\propto \left[\prod\limits_{k=1}^nA({\vec r_{k,l}})\right]^{\dagger}\prod\limits_{k=1}^nA({\vec r}_{k,l}),
\label{prob}
\end{equation}
where the operators $\Pi_{l}$ are semi-positive definite and can be considered as elements of a POVM (positive operator valued measure), while $p_l$ can be considered as the set of targeted probabilities. Generally, $\Pi_{l}$ can be singular and might not form a complete set required for an unambiguous representation of the antenna state. The visibility operator that comprises all possible arrangements of the detectors is $C_V=\sum\limits_{\forall l}\Pi_{l}$ \cite{mog2006}. This operator defines the subspace of states accessible for measurements. The operator $C_V$ can be singular too and different from the unity operator. Thus, an exact solution for the density matrix might not exist for some subsets of targeted probabilities. Therefore, here we consider the problem of shaping the correlation function in the following way: we look for the density matrix (estimator)  maximizing  the probabilities $p_s$ of some subset $\{s\}$, while simultaneously  minimizing other  probabilities $p_m$, $m\in \{l\}\diagdown\{s\}$. Assuming that
\begin{equation}
0\leq p_l^{low}\leq p_l\leq p_l^{high},
\label{cond}
\end{equation}
$p_l^{low(high)}$ being the lower(upper) limit of the targeted probabilities, our design problem can  be formulated as a minimization of some distance between target and estimated sets, $D(p_{target},p_{l})$, where the set of the targeted values is $\{p_m^{low},p_s^{high}\}$. Like the directivity problem for a classical antenna \cite{balan}, the  problem of maximizing the directivity of the quantum antenna can be formulated  in the following way: we look for a conditional minimum of $Tr\{C_V\rho\}$ under the conditions (\ref{cond}) and for $\rho\geq 0$. Note that defining an available target range is a semi-definite programming problem of finding $\min(\max)\{p_l\}$ for $\rho\geq 0$.
As an example, let us take an antenna in the pure two-excitations state: $\rho=|\psi\rangle\langle\psi|$, where
\begin{equation}
|\psi\rangle=\sum\limits_{j=2}^N\sum\limits_{m=1}^{j-1}c_{jm}|+_j,+_m\rangle,
\label{psi1}
\end{equation}
with the summation performed over all distinct pairs of indices $(j,m)\in [1,N]$. The vectors $|+_j,+_m\rangle$ describe the state of excited $j$th and $m$th TLSs with all other TLSs in the lower state. An example of available targeted regions for the state (\ref{psi1}) in array with $N=20$ is shown in Fig. \ref{fig1b}. One can see that limitations on possible choices of targeted probabilities can be quite severe.

\section{Field-state considerations}

To give an intuitive picture of the connection between the state of the antenna and the field correlation functions, let us consider the field of the antenna in the momentum space.
Initially, let us assume that the antenna initial state is a product of the states of the first $M$ fully excited emitters and $N-M$  emitters in the ground state. For time intervals much longer than the inverse decay rate of the excited state $\gamma$, the field disentangles from the emitters and can be written as \cite{scullybook}:
\begin{eqnarray}
|\Psi\rangle \propto \int\prod\limits_{j=1}^M\Bigl[ d^3{\vec k}_ja_j^{\dagger}({\vec k}_j)V({\vec k}_j)\Bigr]\Phi(\{{\vec k}_j\},\{{\vec R}_j\})  |vac\rangle,
\label{psi5}
\end{eqnarray}
where the function
\[
V({\vec k}_j)=\frac{\sqrt{w(\vec{k_j})}{\vec d}{\vec e}(\vec{k_j})}{w(\vec{k}_j)-\omega+i\gamma/2}
\]
does not depend on the positions of the emitters. The function
\begin{equation}
\Phi(\{{\vec k}_j\},\{{\vec R}_j\})=\exp{\{-i\sum\limits_{j=1}^M\vec{k}_j\vec{R}_j\}},
\label{psi2}
\end{equation}
describes the relative phase shifts introduced by the locations of the TLSs and the detectors. Here $a_j^{\dagger}$ is the creation operator for the mode with  momentum $\vec{k}_j$, frequency $w(\vec{k}_j)$ and polarization vector ${\vec e}(\vec{k_j})$; $|vac\rangle$ is the vector of the field vacuum and $\omega$ is the TLS transition frequency. Eqs. (\ref{psi5},\ref{psi2}) give a  hint for understanding the mechanism of $G^{(2)}$ shaping.  Let us take again, for example, the simple two-excitations pure state (\ref{psi1}) with $c_{jm}=\delta_{m,N+1-j}/\sqrt N$. For such an initial state of the antenna, the wave function  of the emitted field state is of the form (\ref{psi5}) with:
\begin{equation}
\begin{gathered}
\Phi(\{{\vec k}_j\},\{{\vec R}_j\})=\frac{1}{\sqrt N}\sum\limits_{m=1}^N\exp{\{-im(\vec{k}_1+\vec{k}_2)\vec{\Delta}\}}
\\
{}\times\exp{\{-i(\vec{k}_1+\vec{k}_2)\vec{R}_0-i\vec{k}_1\vec{\Delta}(N+1)/2\}},
\\
{} = \exp\{-i(\vec k_1 + \vec k_2) \vec R_0\} \frac{\sin\{N(\vec k_1 - \vec k_2) \vec \Delta / 2\}}{\sqrt N \sin\{(\vec k_1 - \vec k_2) \vec \Delta / 2\}},
\end{gathered}
\label{psi3}
\end{equation}
where the vector $\vec{\Delta}=\vec{R}_{m+1}-\vec{R}_{m}$ does not depend on $m$, and $\vec{R}_0$ is the vector describing the position of the antenna middle point. The function $|\Phi|$ in Eq. (\ref{psi3}) for $N \gg 1$ has a sharp peak at $(\vec{k}_1-\vec{k}_2)\vec{\Delta}=0$ and tends towards the delta-function $\delta((\vec{k}_1-\vec{k}_2)\vec{\Delta})$ when $N\rightarrow\infty$. This function is not factorable with respect to momenta $\vec{k}_j$, thus, the state (\ref{psi5}) is entangled in momentum. Hence, one should expect  a sharp maximum in the second-order correlation function $G^{(2)}$, corresponding to co-directionally emitted photons.

Similarly, Eq. (\ref{psi2}) points to the possibility of emitting multi-photons momentum-entangled states and to shape higher-order correlation functions even using the simplest linear array antenna of Fig. \ref{fig1}. Indeed, a superposition of at least two different sets of initially excited antenna TLSs leads to a non-factorability of the function $\Phi(\{{\vec k}_j\},\{{\vec R}_j\})$ and thus to momentum entanglement of the wave-function (\ref{psi5}). Let us demonstrate this effect on the example of a three-photons state with the initial antenna state
$|\psi\rangle = \frac{1}{\sqrt{N-2}}\sum\limits_{j=1}^{N-2}|+_{j},+_{j+1},+_{j+2}\rangle$. We obtain
\begin{equation}
\begin{gathered}
\Phi_l(\{{\vec k}_j\},\{{\vec R}_j\})=2 \exp\{-i(\vec k_1 + \vec k_2 + \vec k_3) \vec R_0\}
\\
{} \times \frac{\sin\{(N-2)(\vec k_1 + \vec k_2+\vec k_3) \vec \Delta / 2\}}{\sqrt {N-2}\,\sin\{(\vec k_1 + \vec k_2 + \vec k_3) \vec \Delta / 2\}}\\
{} \times \Bigl(\cos\{\frac{(\vec k_1 - \vec k_2) \vec \Delta}{2}\}+ \cos\{\frac{(\vec k_1 - \vec k_3) \vec \Delta}{2}\} \\ {} + \cos\{\frac{(\vec k_2 - \vec k_3) \vec \Delta}{2}\} \Bigr)
\end{gathered}
\label{psi4}
\end{equation}
with $|\Phi|$ being not factorable and approaching the delta-function $\delta((\vec{k}_1+\vec{k}_2+\vec{k}_3)\vec{\Delta})$ for $N\rightarrow\infty$. The correlation function  $G^{(3)}$  is sharply peaked for angles satisfying the condition
$(\vec{k}_1+\vec{k}_2+\vec{k}_3)\vec{\Delta}=0$.

\section{Example I: Directional two-photon emission}
\label{examples}

To demonstrate the feasibility of our antenna design approach, we apply it to the simple case of twin-photons generation by the linear  antenna in Fig.\ref{fig1} with initial states (\ref{psi1}). We aim to find the coefficients $c_{jm}$ that provide a desired spatial pattern of the second-order correlation function $G^{(2)}$. Taking into account  considerations from the previous section, we consider the optimization of $G^{(2)}$ for twin-photons emission from a finite length linear array antenna.

\begin{figure}[htb]
\begin{tabular}{cc}
  (a) & (b) \\
  \includegraphics[width=0.45\linewidth]{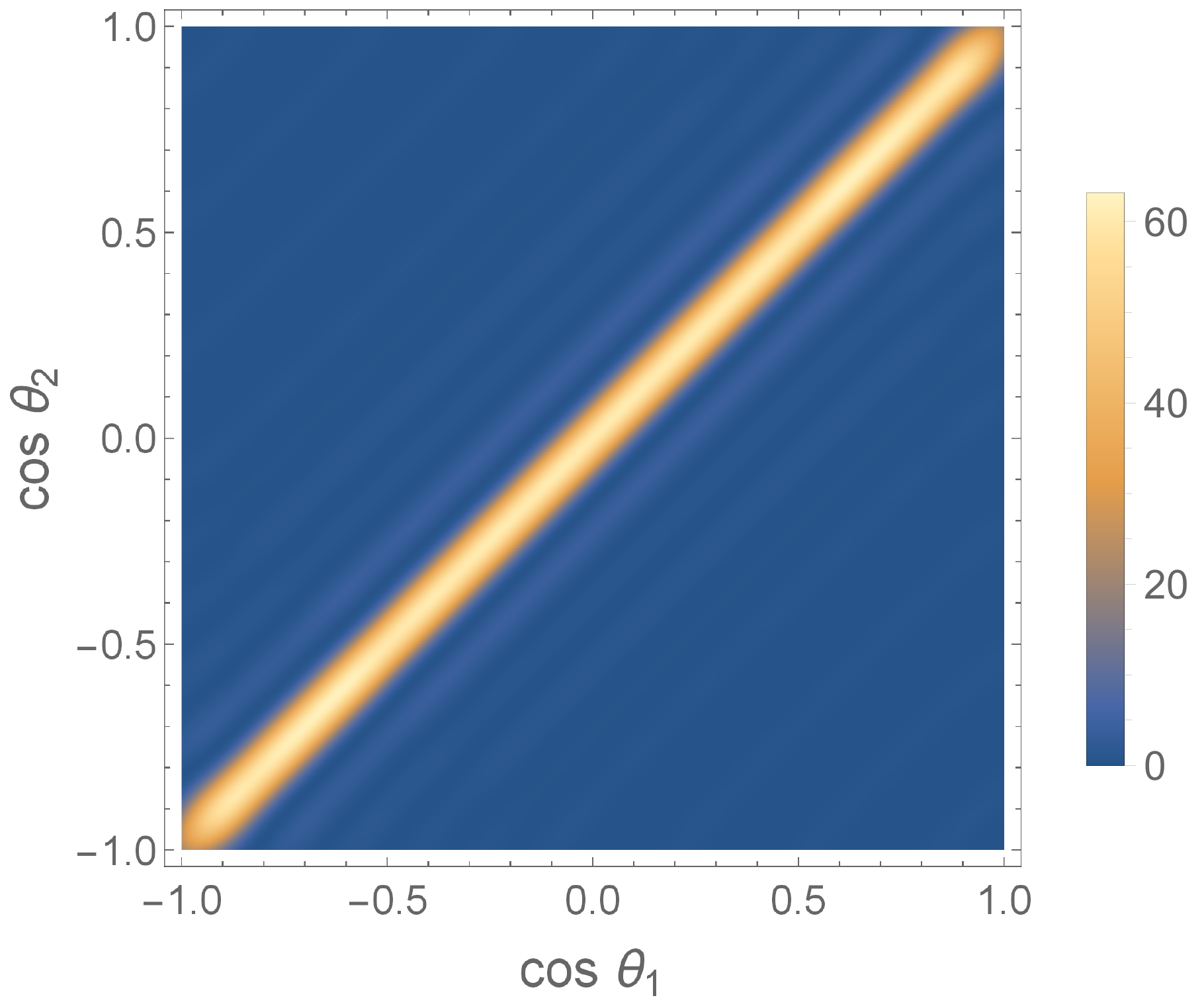} & \includegraphics[width=0.45\linewidth]{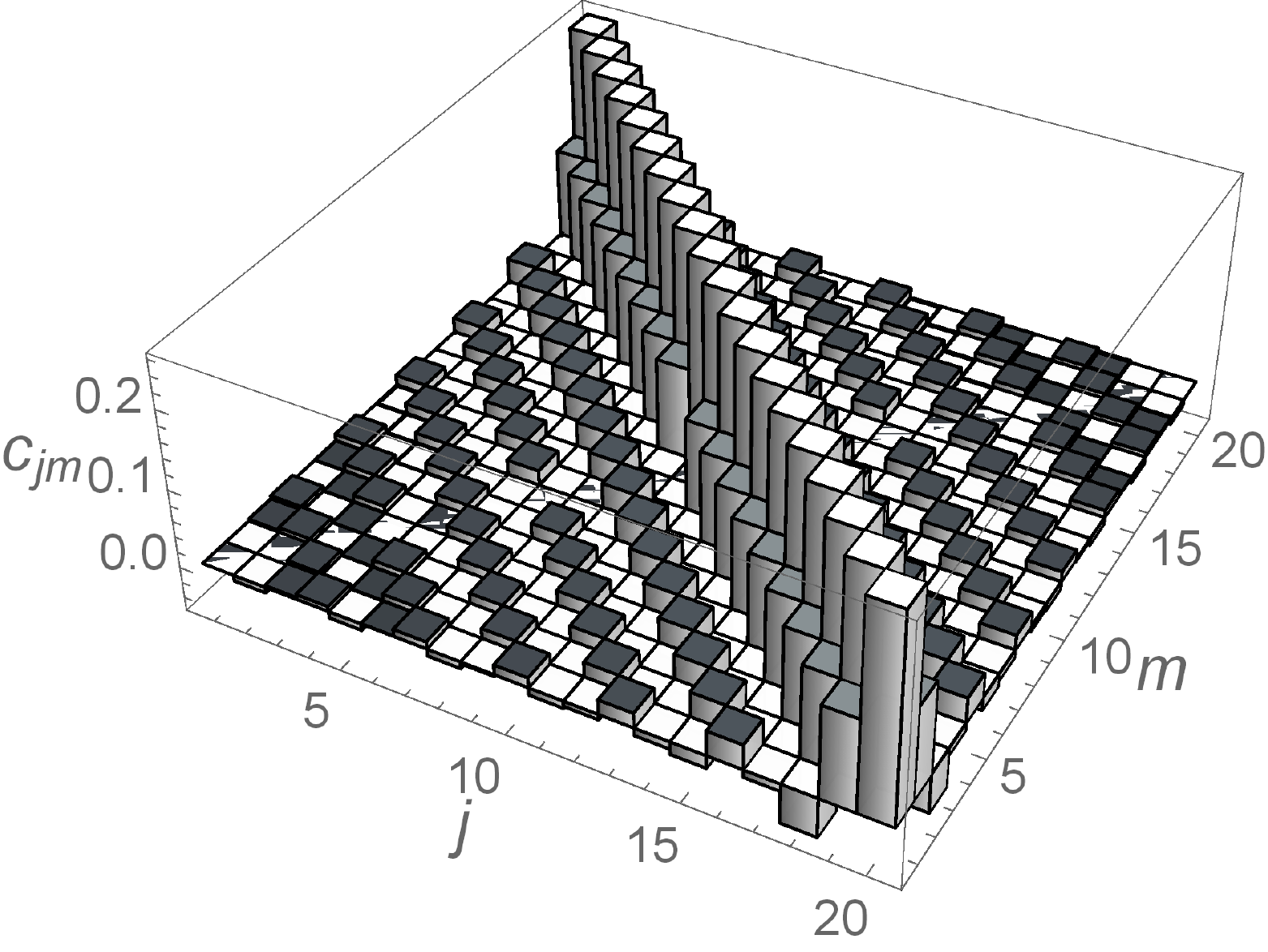} \\
  (c) & (d) \\
  \includegraphics[width=0.45\linewidth]{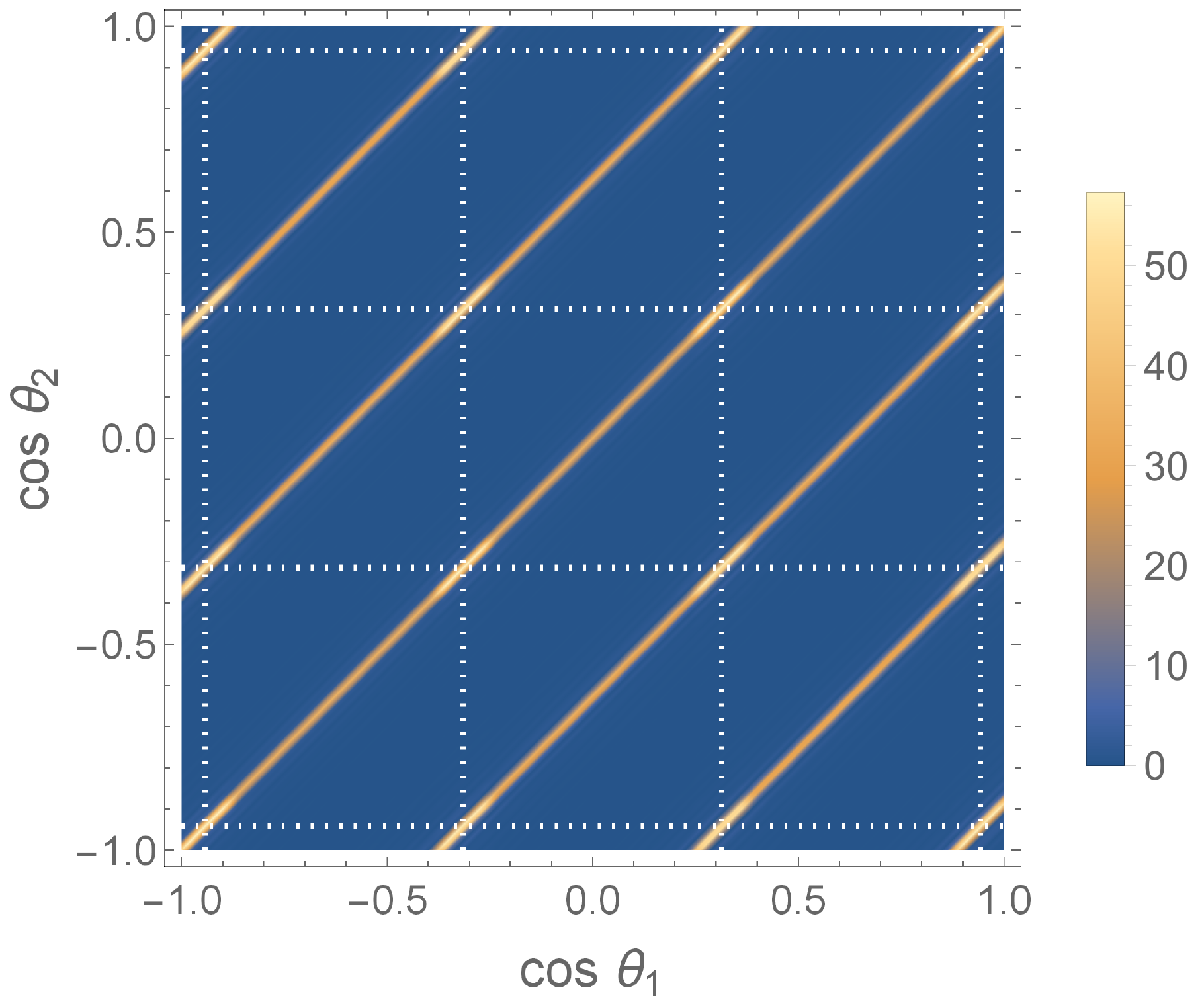} & \includegraphics[width=0.45\linewidth]{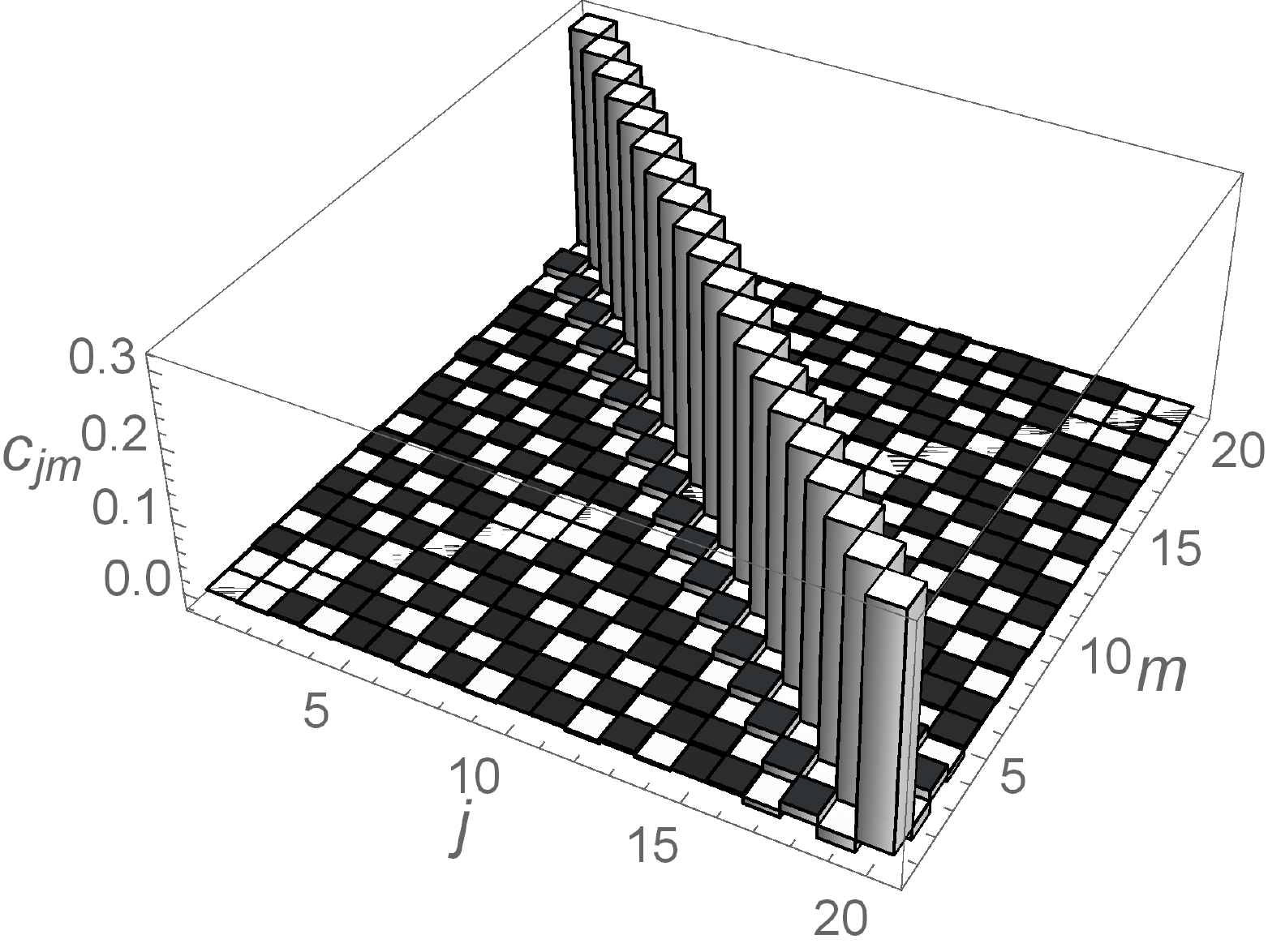} \\
\end{tabular}
\caption{(a), (c) The normalized  $G^{(2)}(\theta_1,\theta_2)$  (\textit{i.e.} $p(\theta_1,\theta_2)$ of Eq. (\ref{g2})) obtained as the result of optimization for the generation of two co-directional photons; (b), (d) states in Eq. (\ref{psi1}) obtained as the results of  optimization. For all panels, $N=20$; for panel (a,b) $k \Delta = 2$, for panels (c,d) $k \Delta = 10$. Dashed lines divide the plot in panel (c) into equivalent regions due to the periodicity of the signal for $k \Delta > \pi$.} \label{fig2}
\end{figure}

\paragraph{Co-directional two-photons emission.}
From Eqs. (\ref{arop},\ref{g}) the second-order correlation function in the plane perpendicular to the orientation of the dipoles is given by
\begin{equation}
\label{g2}
\begin{gathered}
G^{(2)}(\theta_1,\theta_2)\propto p(\theta_1, \theta_2) =Tr\{\Pi(\theta_1,\theta_2)\rho\}, \\
\Pi(\theta_1,\theta_2)=\sum\limits_{j,m,n,q}\exp\{ik\Delta(j-n)\cos(\theta_1)\}\times\\
\exp\{ik\Delta(m-q)\cos(\theta_2)\}\sigma_{j}^+\sigma_{m}^+\sigma_{n}^{-}\sigma_{q}^{-},
\end{gathered}
\end{equation}
where $k$ is the wave-number and $\Delta$ is the distance between the dipoles; $\theta_{1,2}$ are the angles in the direction of the detectors. Optimization of the antenna directivity for this case can be formulated as a quadratic programming problem of minimizing the average visibility operator $\langle\psi|C_V|\psi\rangle$ subjected to conditions (\ref{cond}). Let us aim, for example, to obtain the co-directional correlation of emitted photons, \textit{i.e.}  sharply peaked $G^{(2)}$ pattern for $\theta_1=\theta_2$. Fig. \ref{fig2}(a,b) shows the results of such optimization for $k \Delta = 2$. The optimization was done by minimizing the weighted sum of the average visibility operator $\langle\psi|C_V|\psi\rangle$ and the quadratic distance between the actual values of $p(\theta_i, \theta_i; c_{jm})$ as well as the targeted value $p_0$ for 100 discrete angles $\theta_i$ in the range $[0, \pi]$. For the  TLS number $N$ varying from 2 to 10, the problem was solved for the general case of complex coefficients $c_{jm} \in \mathbb{C}$. However, the  imaginary parts of the optimal solution turned out to be very small in comparison with the real parts of $c_{jm}$. Therefore, for antennas with a larger number of TLSs (we checked up to $N = 20$), these coefficients were assumed to be real: $c_{jm} \in \mathbb{R}$. Indeed, in Fig. \ref{fig2}(a) one can see that $G^{(2)}$ is sharply peaked around equal observation angles. The initial antenna state  producing such correlations is shown in Fig. \ref{fig2}(b). Remarkably, in accordance with the field-state considerations of the previous section, we have obtained that $c_{j,N+1-j}\approx1/\sqrt N$, while all other coefficients are much smaller. The initial state corresponds to excited pairs of the dipoles  located symmetrically on the opposite sides  of antenna (\textit{e.g.}, the first and the last one, the second and the $(N-1)$-th, etc.).  Fig. \ref{fig2}(c,d) shows the optimization results for larger distance between dipoles, \textit{i.e.}, $k \Delta = 10$. The radiation pattern is still sharply peaked around $\theta_1 = \theta_2$, but in contrast to the previous example, each emission direction of one photon is correlated to several possible emission directions of the second photon. To elucidate the origin of this pattern, we rewrite Eq. (\ref{g2}) as:
\begin{equation}
\begin{gathered}
p(\theta_1, \theta_2) =  \langle \psi | \Pi(\theta_1, \theta_2)|\psi\rangle = |\Phi(\theta_1,\theta_2)|^2 \\ {} \equiv \left|\sum_{j,m=1}^N c_{jm} \exp\{-ik\Delta(j \cos\theta_1 + m \cos\theta_2)\} \right|^2.
\end{gathered}
\label{g2pure}
\end{equation}
Indeed, it can be seen that $\Phi(\theta_1,\theta_2)$ is a periodic function of $\cos \theta_1$ and $\cos \theta_2$, that is $\Phi(\theta_1, \theta_2) = \Phi(\theta_1, \theta_2')$ if $k \Delta (\cos \theta_2 - \cos \theta_2') = 2 \pi n$ with integer $n$.

\begin{figure}[htb]
\begin{tabular}{cc}
  (a) & (b) \\
  \includegraphics[width=0.45\linewidth]{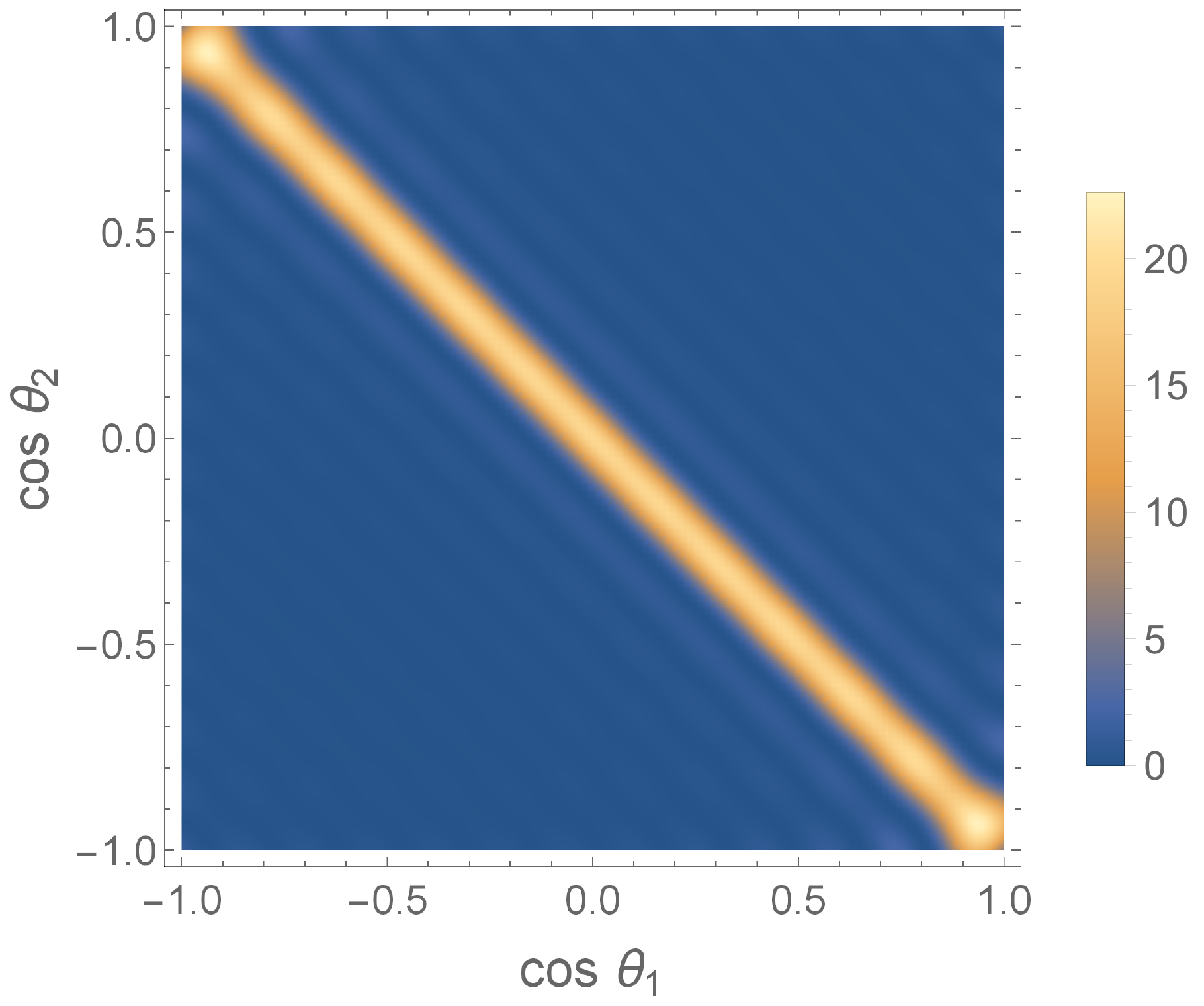} & \includegraphics[width=0.45\linewidth]{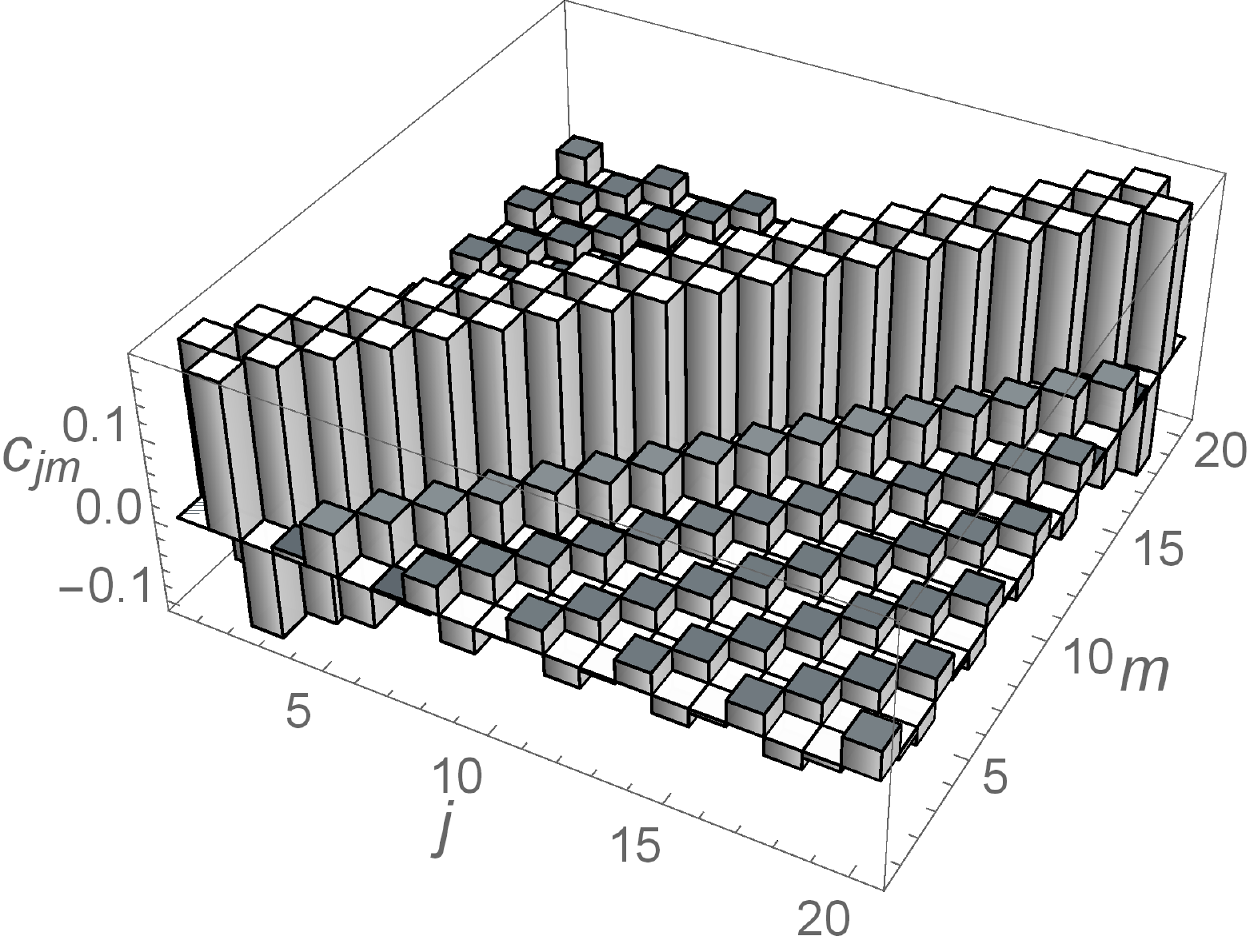} \\
  (c) & (d) \\
  \includegraphics[width=0.45\linewidth]{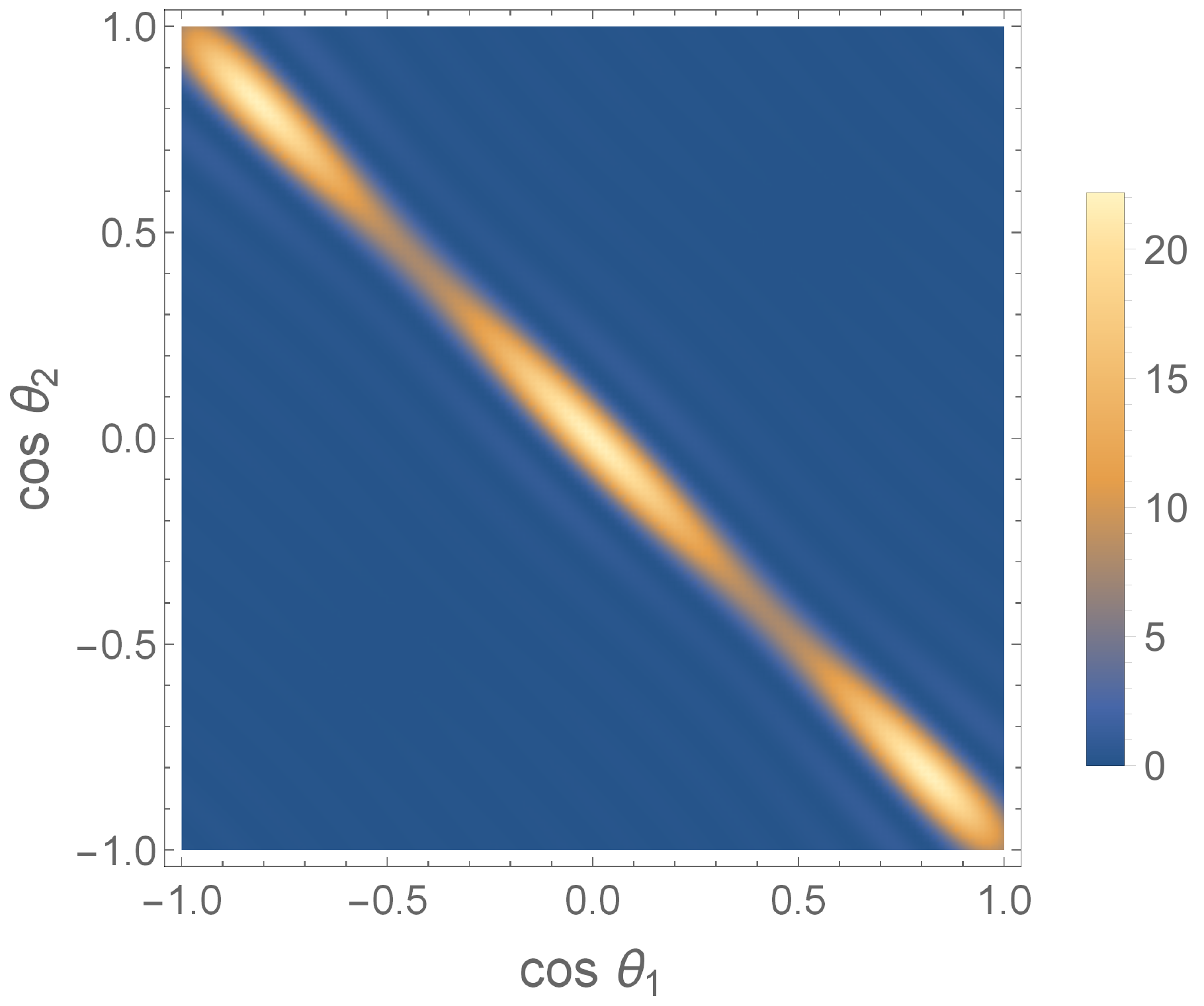} & \includegraphics[width=0.45\linewidth]{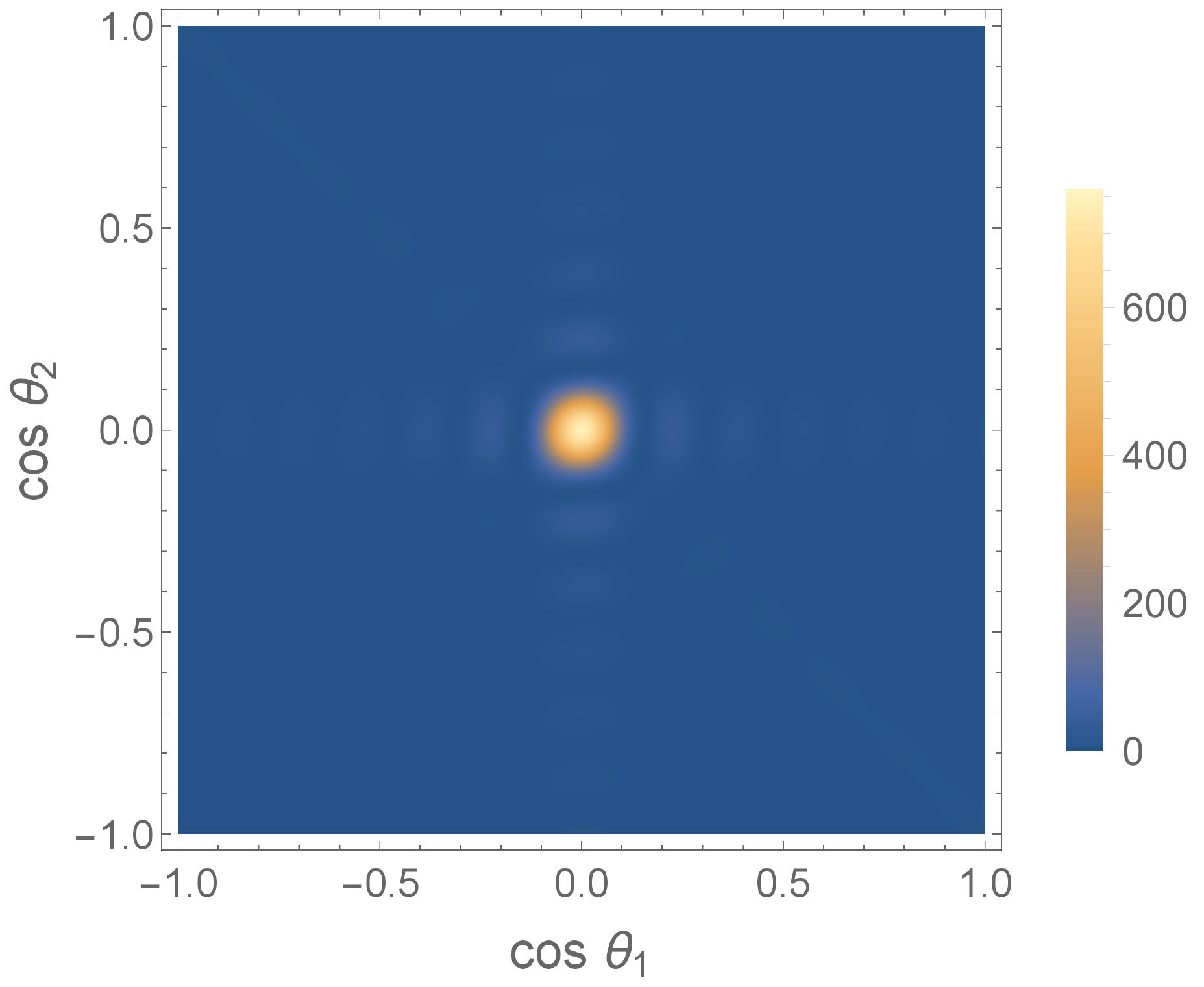} \\
\end{tabular}
\caption{(a), (c) The normalized $G^{(2)}(\theta_1,\theta_2)$ obtained as the result of numerical and analytical optimization for the generation of two contra-directional photons, respectively; (b) the state in Eq. (\ref{psi1}) obtained as the result of numerical directivity optimization. (d) The normalized $G^{(2)}(\theta_1,\theta_2)$ obtained as the result of optimization for the generation of two photons emitted in a direction perpendicular to the antenna. For all panels, $N=20$; $k \Delta = 2$.} \label{fig3}
\end{figure}

\paragraph{Contra-directional two-photons emission.}
 By tailoring the initial quantum antenna state, one can also achieve contra-directional correlations between emitted photons. The optimization results for the radiation pattern with $G^{(2)}$ sharply peaked around $\theta_2 = \pi - \theta_1$  are shown in Fig.~\ref{fig3}(a), which attests  strong contra-directional correlations. The optimal initial state of the antenna (Fig.~\ref{fig3}(b)) shows a peculiar structure of the  matrix $c_{jl}$ describing the state (\ref{psi1}). This matrix is composed of sets of the coefficients  with equal amplitudes on each sub-diagonal, \textit{i.e.} coefficients $c_{j,j\pm l} = c_l$ do not depend on the index $j$. Once again, this feature can also be  explained using field-state considerations on the basis of Eqs.(\ref{psi5},\ref{psi2}) in the following way. Let us consider the contribution from just one sub-diagonal of the matrix $c_{j,j+l}$ with index shift  $l$ (\textit{i.e.} we assume  $c_l = 1/ \sqrt{N-l}$ and  $c_{l'}=0$ for all other sub-diagonals with $l'\ne l$). The wave function of the emitted field state is described by Eq.(\ref{psi5}) with:
\begin{equation}
\begin{gathered}
\Phi_l(\{{\vec k}_j\},\{{\vec R}_j\})=2 \exp\{-i(\vec k_1 + \vec k_2) \vec R_0\}
\\
{} \times \cos\{\frac{l}{2}(\vec k_1 - \vec k_2) \vec \Delta\} \frac{\sin\{(N-l)(\vec k_1 + \vec k_2) \vec \Delta / 2\}}{\sqrt {N-l}\,\sin\{(\vec k_1 + \vec k_2) \vec \Delta / 2\}}.
\end{gathered}
\label{psi3b}
\end{equation}
For $N\rightarrow \infty$ and finite $l$ the function $|\Phi_l|$ asymptotically tends toward the delta-function $\delta ((\vec k_1 + \vec k_2)\vec \Delta)$, which corresponds to an entangled two-photons state with strong contra-directional correlations. However, in contrast to the previous example in sub-Section V.a, the absolute value of the wave function is varied along the line $\vec k_2 \vec \Delta = - \vec k_1 \vec \Delta$ ($\theta_2 = \pi - \theta_1$) as $\cos (l k \Delta \cos \theta_1)$. In order to obtain a $G^{(2)}$ pattern with even contra-directional correlations as shown in Fig.~\ref{fig3}(a), one needs to combine several sub-diagonal sets $c_{j,j\pm l} = c_l$ with different sub-diagonal numbers $l$. Fig.~\ref{fig3}(c) shows the result of such a combination for three sets with $l=1$, 2, 3 and relative amplitudes $c_1 : c_2 : c_3 = 1 : {-0.7} : 0.4$. The state shows strong contra-directional correlations across the full range of angles (see the main diagonal in the pattern of Fig.~\ref{fig3}(c)), but still it gives  a less even and less sharply directed pattern of $G^{(2)}$ than the state  found by numerical optimization in Fig.~\ref{fig3}(b).

\paragraph{Maximal directivity of emission.}
As a particular example one can consider a state with the maximal directivity of two-photons emission in the direction perpendicular to the linear array antenna ($\theta_1 = \theta_2 = \pi / 2$). The radiation pattern for the numerically optimized state is shown in Fig.~\ref{fig3}(d). As one would expect, the optimal state is close to the symmetric two-excitations Dicke state with $c_{jm} = const$ for all indexes $j$ and $m$.

\section{Example II: "Dark" states and antenna design}

\begin{figure}[htb]
\begin{tabular}{cc}
  (a) & (b) \\
  \includegraphics[width=0.45\linewidth]{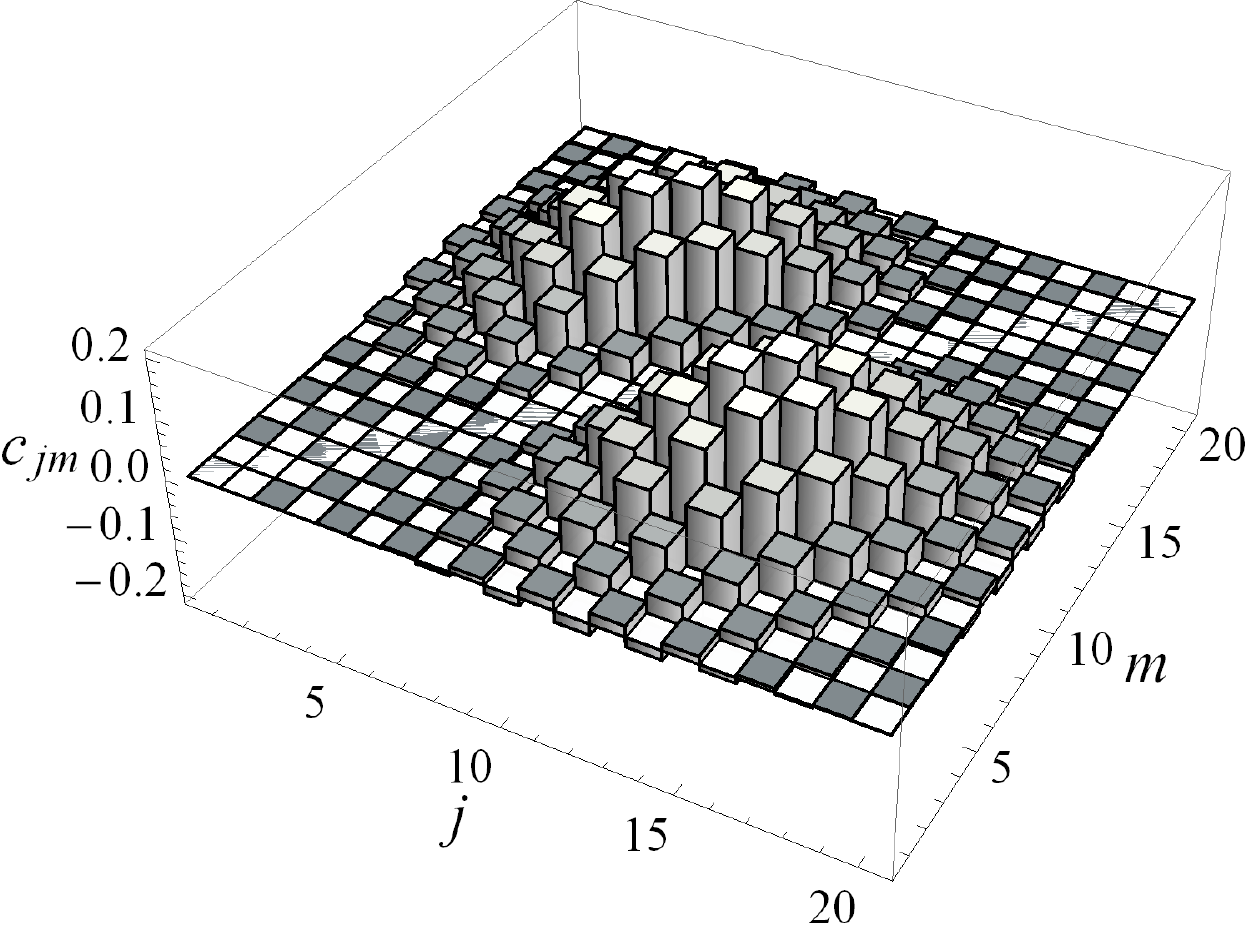} & \includegraphics[width=0.45\linewidth]{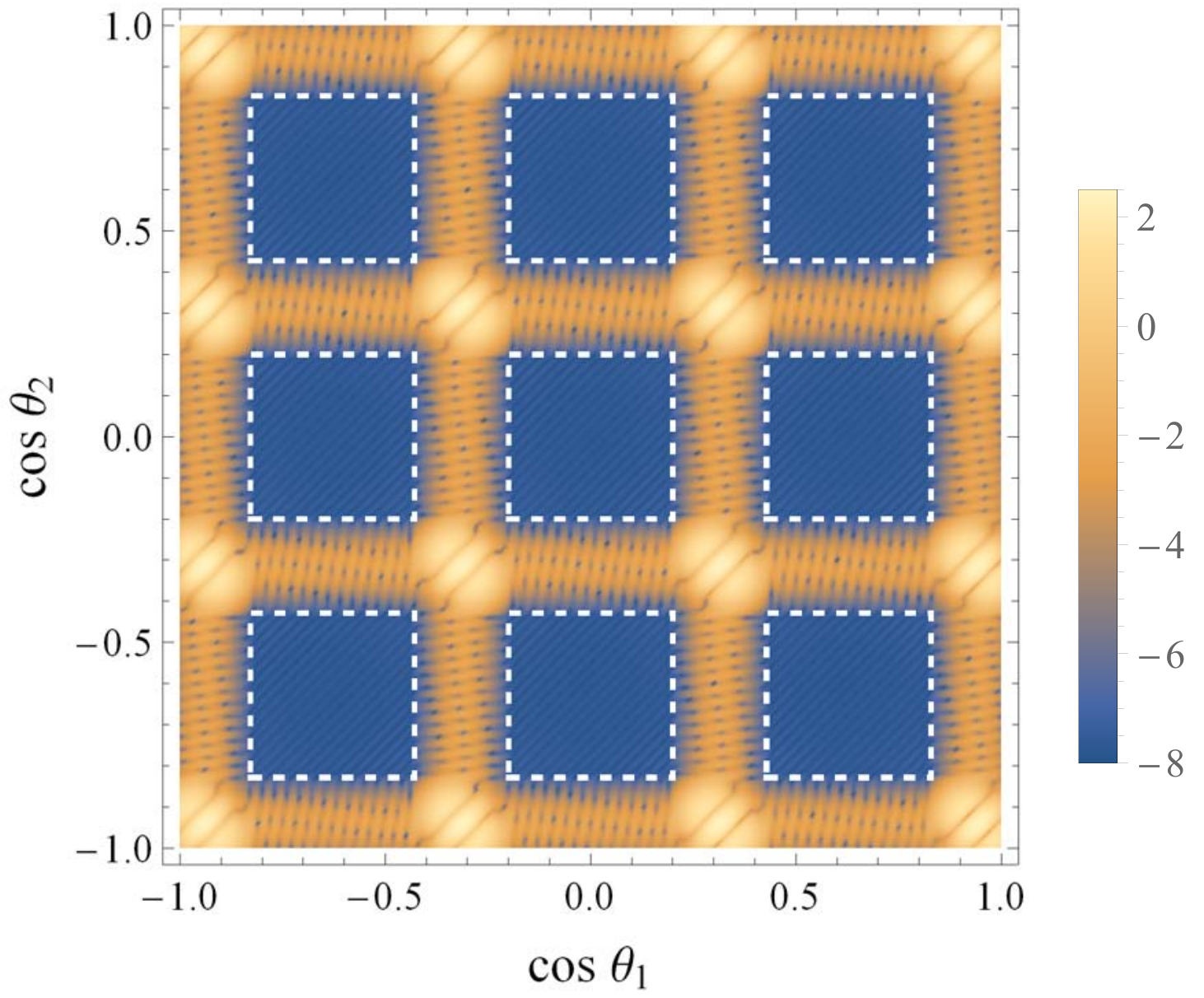}
\end{tabular}
\caption{(a) The "dark" state (\ref{psi1}) obtained as the result of the optimization for $N = 20$, $k\Delta = 2$. (b) The pattern of $\log_{10} G^{(2)}(\theta_1,\theta_2)$ for the state (a), calculated for $k \Delta = 10$. Dashed lines define "dark" regions.} \label{fig4}
\end{figure}

 \paragraph{Finding the "dark state".}
Localization of the emitted field inside a finite volume is something that one would really expect for such exquisitely designed objects as 3D photonic crystals and meta-material structures  \cite{gaponenko,shalaev}. With classical antennas, one can specifically design such distributions of currents so as to obtain the same effect, that is to create a ``non-radiative source'' \cite{balan,wolf}.
 Counterintuitively, this effect can also be achieved in a simple regular linear antenna array by choosing the appropriate initial quantum state of the antenna.
  Just by minimizing $G^{(2)}(\theta_1,\theta_2)$ for all angles $\theta_1$ and $\theta_2$, one obtains the initial state of the antenna leading to a strong field suppression in the far-field zone. Indeed, let us minimize the average visibility operator $\langle\psi|C_V|\psi\rangle$ for the state (\ref{psi1}) without imposing any additional requirements to obtain bright spots or lines in the $G^{(2)}$ patterns. For $k\Delta < \pi$, such optimization can be successfully performed (Fig.~\ref{fig4}(a)), yielding the maximum  $G^{(2)}$ value in Eq. (\ref{g2})   of $3\cdot 10^{-7}$  for $N=20$ and $k\Delta = 2$. Eq.~(\ref{psi3b}) gives a hint on how to design such a state: first, choose broad distributions of the matrix coefficients describing the state (\ref{psi1}) along each sub-diagonal, \textit{i.e.}  take $c_{j,j\pm l} \propto f(j - (N+1)/2)$ along sub-diagonals to suppress emission outside the region with $\theta_2 \approx \pi - \theta_1$. For example, a Gaussian distribution of the matrix coefficients $f(m)=e^{-m^2/\sigma^2}$ leads to $k\Delta|\cos\theta_1 + \cos \theta_2| \lesssim 2 / \sigma$. Then, one should suppress the emission along the diagonal using an appropriate combination of $\cos\{l k \Delta (\cos \theta_1 - \cos \theta_2) / 2\}$ (see Eq.~(\ref{psi3b})). Here, one can find an approximate analytical expression surprisingly close to the  optimal state numerically found  in Fig.~\ref{fig4}(a), \textit{i.e.}:
\begin{equation}
c_{jm}\propto (-1)^l l^2 \exp\{-(l^2+q^2) / (4 \sigma^2)\},
\label{dark}
\end{equation}
where $l = |j - m|$, $q = (j + m) - (N + 1)$, and $\sigma \approx 3.2$.
It is worth mentioning that for $k\Delta > \pi$ one cannot design such a ``dark" state. By introducing dimensionless variables $x_j = k\Delta \cos \theta_j$, $j = 1,2$, $x_j \in [-k\Delta, k\Delta] \supset [-\pi, \pi]$, one can easily see that the  following lower bound holds:
\begin{equation}
\begin{gathered}
\int\limits_{-k\Delta}^{k\Delta} dx_1 \int\limits_{-k\Delta}^{k\Delta} dx_2 p(\theta_1, \theta_2) \ge \int\limits_{-\pi}^\pi dx_1 \int\limits_{-\pi}^\pi dx_2 p(\theta_1, \theta_2) \\
{} = 2 \pi^2 \sum_{j=2}^{N} \sum_{m=1}^{j-1} |c_{jm}|^2 = 2 \pi^2,
\end{gathered}
\label{norm}
\end{equation}
where the function $p$ is defined by Eq.~(\ref{g2}) and the normalization of the state (\ref{psi1}) is taken into account. Fig.~\ref{fig4}(b) shows the radiation pattern for $k \Delta = 10$ while the initial state is depicted in Fig.~\ref{fig4}(a). One can see  in Fig.~\ref{fig4}(b) that  it is possible to suppress emission in some regions (blue squares with dashed border in Fig.~\ref{fig4}(b)), but not to the whole range of angles. Note that just one square would represent  the total radiation pattern for $k \Delta = 2$.

Remarkably, by unconditionally maximizing the mean value of the visibility operator $\langle\psi|C_V|\psi\rangle$ for the state (\ref{psi1}), we can achieve an opposite effect and obtain a nearly homogeneous far-field distribution.

\paragraph{Simultaneous state and antenna design.}
Interestingly, the field-state considerations also point to the possibility of engineering the initial state and the antenna geometry for a complete  3D suppression of the far field. Indeed, let us take two perpendicular linear antennas with the same dipole moments orthogonal to the antenna plane and randomly located TLSs. We choose the initial state as a superposition of randomly chosen pairs of TLS from the first and second antennas. Then, the phase factor reads
\begin{equation}
\Phi(\{{\vec k}_j\},\{{\vec R}_j\})\propto\sum_{\forall j,m}\exp{\{-i(\vec{k}_1\vec{R}_j+\vec{k}_2\vec{R}_m)\}},
\label{psi7}
\end{equation}
where  $j$ and $m$ are the indices of the TLSs from the two antenna arms. For a sufficiently large number of TLSs in the antenna, the phase factor in Eq. (\ref{psi7}) tends toward zero for all directions $\vec{k}_{1,2}$ except for directions parallel to $\vec{d}$. In this way, the emitted field in the far-field zone is suppressed.
However, one should notice that for the states predicting a field localization, the antenna approximation of non-interacting emitters might fail. The localized photons might be re-absorbed and re-emitted by the antenna (in section VII below we outline one possible approach to account for such interactions between emitters) .

 Also, by simultaneously changing the shape and the initial state of the antenna, one can get a high directivity of the correlation function without using the Dicke state as the initial antenna state. Eq. (\ref{psi3}) hints toward a simple way to obtain co- or contra-directional correlations of emitted photons that are localized in a narrow region in the vicinity of $\pm\pi/2$. More specifically, instead of one regular antenna array shown in Fig. \ref{fig1}(a), let us consider an antenna composed of two regular linear arrays located on the same axis and each comprising $N$ TLSs. We choose the pitch of TLSs in one sub-antenna to be $u$-times larger than in  the other sub-antenna.

We consider the initial antenna  state (\ref{psi1}) with excited TLS pairs composed of one counterpart from the first sub-array (\textit{e.g.} with larger pitch) and another counterpart from the second sub-array (\textit{e.g.} with smaller pitch). In this case the first index of the matrix element $c_{jm}$  enumerates TLS from the sub-array with larger pitch while the second index enumerates TLS from the the sub-array with shorter  pitch. As in the section V.a, we then additionally impose a specific symmetry between indexes of TLS  located symmetrically on the opposite sides  of antenna arms. More specifically, we
define the initial state by non-vanishing coefficients $c_{j,N+1-j}=1/\sqrt N$, where the index $j$ spans over the large-period sub-array (the index $N+1-j$ then spans over a short-period sub-array). For our compound antenna composed of two sub-arrays with different pitches, Eqs. (\ref{psi2},\ref{psi3}) suggest a sharp localization of the emitted photons for $(\vec{k}_1-\vec{k}_2/u)\vec{\Delta}=0$, which can be satisfied for $u\gg1$ only if both $\vec{k}_{1,2}$ are nearly orthogonal to $\vec{\Delta}$.

Thus, we can see that a simultaneous design of the antenna geometry and the initial states opens considerably richer possibilities for the optimization of the correlation functions compared to the state design for a pre-defined antenna. However, generally, such a design is a complicated  nonlinear optimization problem.

\begin{figure}[htb]
   \includegraphics[width=0.5\linewidth]{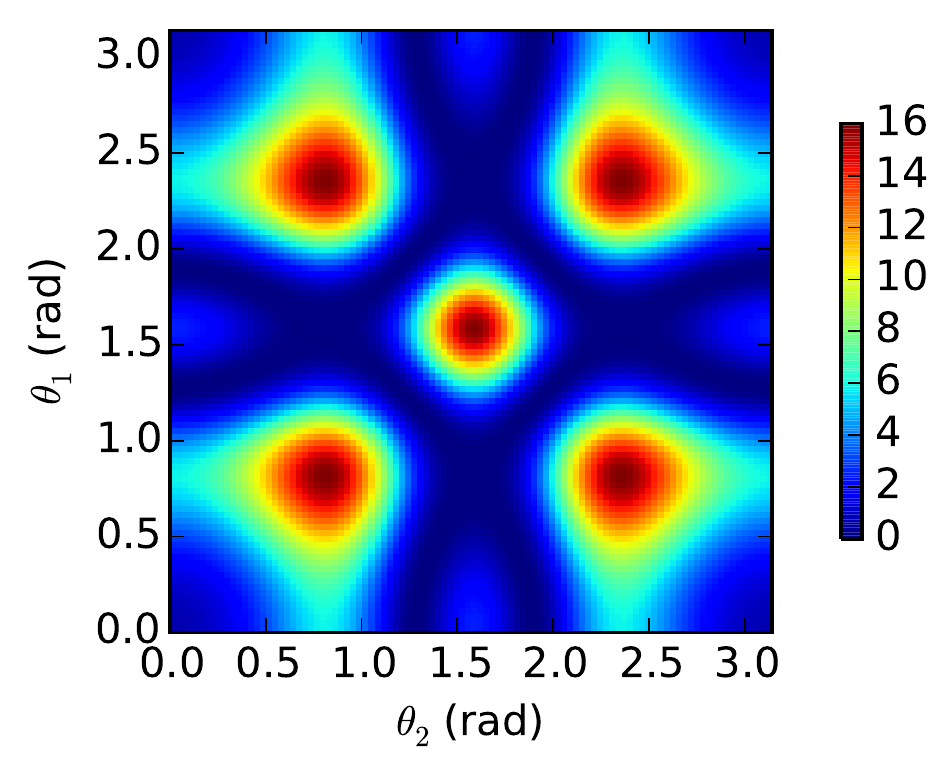} \\
\caption{The normalized quantum correlation function $G^{(2)}(\theta_1,\theta_2)$ for the initial state with $N=3$ giving
contra-directional twin-photons correlations. The dipole-dipole distance is $k\Delta=4.5$. }
\label{fig5a}
\end{figure}

\begin{figure}[htb]
 \begin{tabular}{cc}
   (a) &  (b) \\
    \includegraphics[width=0.51\linewidth]{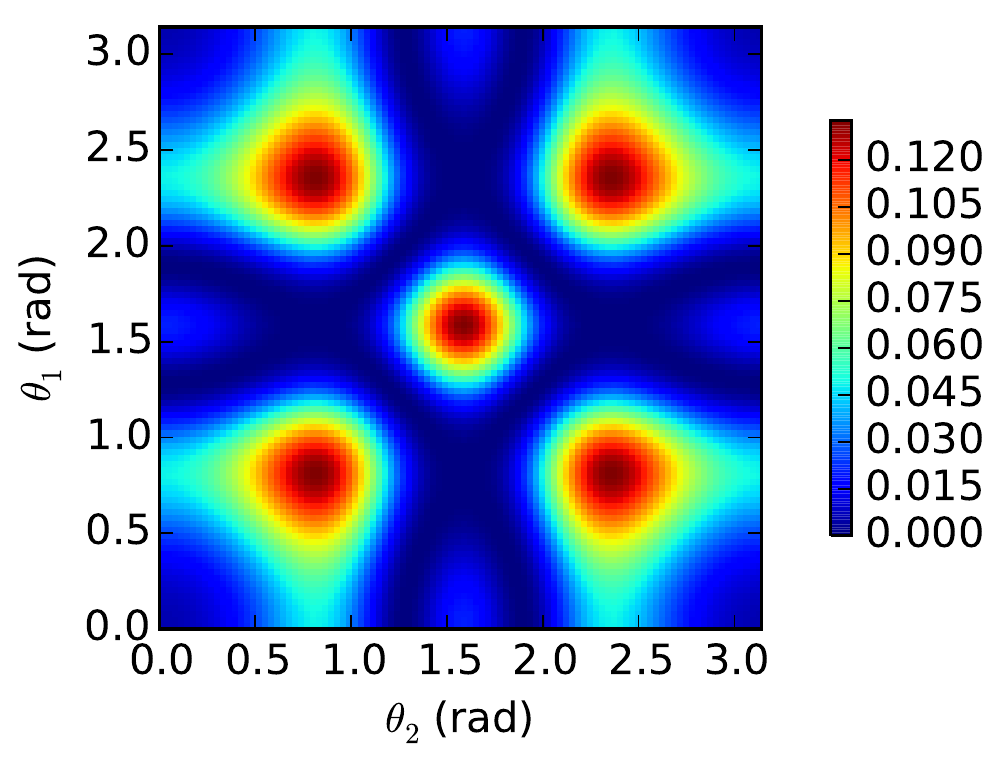} &  \includegraphics[width=0.5\linewidth]{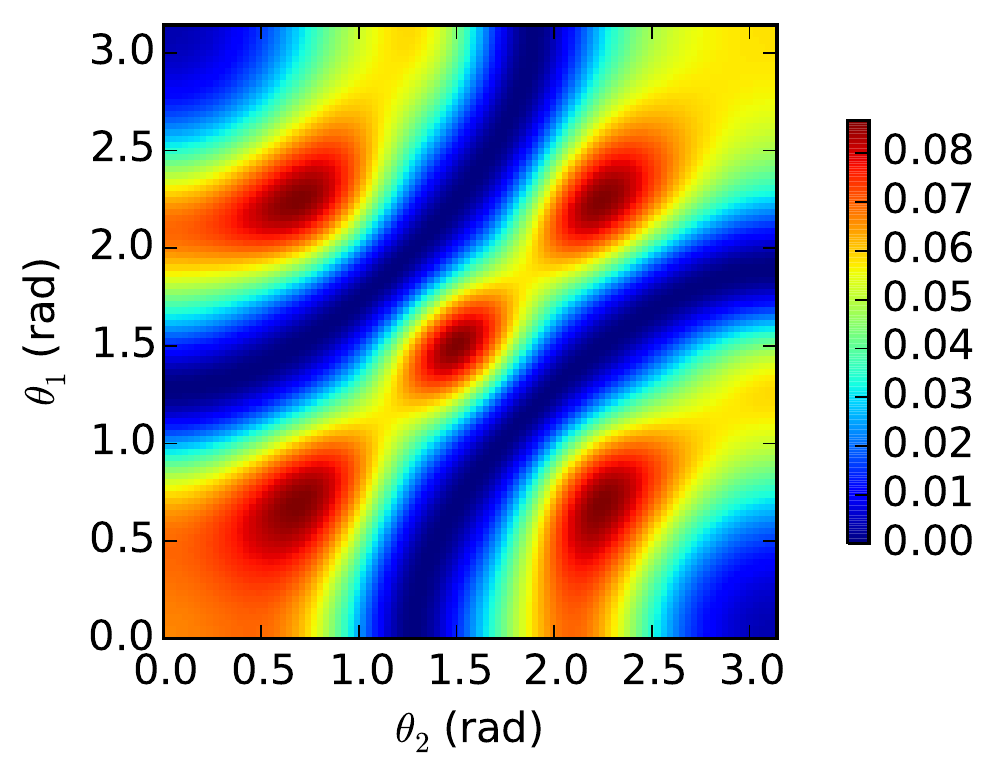}
    \end{tabular}
\caption{The nornalized semiclassical correlations functions $G^{(2)}(\theta_1,\theta_2)$ for the initial state with $N=3$ giving
contra-directional twin-photons correlations. The semiclassical solution is given by Eq.~(\ref{gp}) and Eqs.~(17) in the Appendix.
  Panel (a) displays the post-semiclassical solution for correlated, "mirrored" noise sources. Panel (b) displays the semiclassical solution for uncorrelated noise sources. The patterns are taken at a specific time and are averaged over 100 realizations (see Appendix). Other parameters are as for Fig.~\ref{fig5a}. }
\label{fig5b}
\end{figure}

\section{Semiclassical approach}

As we have shown, quantum interferences are essential for shaping the correlation functions. Here we show that it is still possible to use a semiclassical approach for modelling the emitters dynamics and, after a minor modification,  to reproduce non-classical features of the spatial correlation functions $G^{(n)}$ from Eq. (\ref{g})  (we have termed this recipe ``the post-semiclassical approximation" ).
Such a recipe can be developed inspite of the fact that the semiclassical approach is, generally,  unable to capture the mechanism of spontaneous emission and effects stemming from it. For example, creation of entanglement between TLS decaying into the same radiative reservoir \cite{ficek} can hardly be captured by the approach assuming an absence of quantum correlations between TLSs.  Nevertheless, field correlation effects can still be successfully captured in some cases.  The best known example is superradiance. The onset of cooperative effects and phase correlations leading to the formation of the superradiant field  pulse  can be quite accuratly described by the semiclassical approach \cite{boiko}.  Importantly, the initial state of interacting semiclassical  TLS does not need to be correlated. The correlations self-establish at the initial stage of cooperative emission \cite{benedict}.

The key observation enabling to develop a post-semiclassical approximation, is given by Eqs. (\ref{arop},\ref{g}).  They show that by modeling  the TLS correlation functions with enough accuracy, one would get an accurate description of the emitted field in the far-field zone.  From the first glance, to model correlation functions of non-interacting quantum emitters with correlation functions of interacting semiclassical emiters (notice that  interaction is intrinsic for semiclassical models) is hardly possible. However, it was shown that interacting TLS can have spatial correlation functions coinciding with the spatial correlation functions of non-interacting TLS \cite{ficek0}. The task is considerably simplified by requiring closeness of only the spatial correlation patterns for some specific time-intervals.

 The semiclassical approach in its simplest form assumes a factorization of the correlation functions (\ref{g}) up to the first-order averages, for example, $\langle\sigma^-_j(t_1)\sigma^-_k(t_2)\rangle\approx\langle\sigma^-_j(t_1)\rangle\langle\sigma^-_k(t_2)\rangle$
where the time-dependence of averages, $\langle\sigma_j^{\pm}\rangle$, is derived from the semiclassical Maxwell-Bloch equations \cite{benedict,boiko} (see also Appendix) .
Firstly, let us consider  semiclassically the antenna with uncorrelated identical initial states of each TLS. Assuming interacting TLSs, for a sufficiently long antenna ($N\gg1$) one can, \textit{e.g.},  replace the sum
$\sum \Pi^{jm}_{nq}(\theta_1,\theta_2)\langle\sigma_{j}^+\sigma_{m}^+\sigma_{n}^{-}\sigma_{q}^{-}\rangle$ in Eqs. (\ref{g},\ref{prob}) with
$\sum \Pi^{jm}_{nq}(\theta_1,\theta_2)\langle\sigma_{j}^+\rangle\langle\sigma_{m}^+\rangle\langle\sigma_{n}^{-}\rangle\langle\sigma_{q}^{-}\rangle$, and finally obtain $\sum \Pi^{jm}_{nq}(\theta_1,\theta_2)|\langle\sigma\rangle|^4$ for the approximation of the correlation function within the relative accuracy   of the order of $N^{-2}$, which is essentially a classical radiation pattern \cite{benedict}.
However, one can extend the semiclassical approach of an interacting TLS antenna for the consideration of
a non-interacting quantum TLS antenna by accounting for commutation relations of TLS operators and correlations between initial state components ( it is the backbone of the recipe for ``the post-semiclassical approximation").

Let us illustrate this concept with our example of  two-excitations initial state giving contra-directional correlations, $c_{jm}=\delta_{j,j+1}/\sqrt{N}$ for the state (\ref{psi1}). For the quantum antenna of non-interacting TLSs the simultaneous second-order correlation function in the far-field zone for the two-excitations state (\ref{psi1}) reads as \cite{scullybook}:
$G^{(2)}(\theta_1,\theta_2;t)=|R(\theta_1,\theta_2,t)|^2$, with
\begin{eqnarray}
\label{gp}
R(\theta_1,\theta_2,t)=\langle vac|\langle -|E(\theta_1,t)E(\theta_2,t)|\psi\rangle|vac\rangle\propto \\
\nonumber
\sum\limits_{j=1}^{N-1}\sum\limits_{l=1,2}\exp\{i\phi_j(\theta_l)+i\phi_{j+1}(\theta_{3-l})\}\\
\nonumber
\langle -|\sigma_j^-(t)\sigma_{j+1}^-(t)|\psi\rangle,
\end{eqnarray}
where $E(\theta_1,t)$ is the field operator, $\phi_j(\theta_l)=k\Delta\cos\theta_l$ and $\langle -|$ is the bra-vector denoting the ground state of all TLSs.
We aim to estimate Eq. (\ref{gp}) semiclassically.
Our recipe for this case would be to consider  the antenna with interacting TLSs semiclassically for different uncorrelated initial states with a pair of neighbour TLS in the excited state and others in the ground state, such as, \textit{e.g.}, $|+\rangle_1|+\rangle_2\prod\limits_{j=3}^N|-\rangle_j$. Then, we would sum the results for all the initial states with phase factors given by Eq.(\ref{gp}) , replacing  $\langle -|\sigma_j^-(t)\sigma_{j+1}^-(t)|\psi\rangle$ with $\langle\sigma_j^-(t)\rangle\langle\sigma_{j+1}^-(t)\rangle$.
Note that the radiation in the semiclassical approach is assumed to be initiated by a random polarization noise source.
Such an approach can lead to spatial patterns of the semiclassical correlation functions which are quite close to the quantum ones even for a small number of TLSs in the antenna. This holds  under the condition of a specifically correlated noise for different initial states (different states with $c_{jm} \neq 0$ in the superposition state  \ref{psi1})   with the aim to reproduce the phase relationships between the parts of the initial superposition state. The details of the semiclassical approach are described in the Appendix.

Fig. \ref{fig5a} shows an example of a quantum pattern of $G^{(2)}(\theta_1,\theta_2)$ for $N=3$ and the initial state $|\psi\rangle\propto|+\rangle_1|+\rangle_2|-\rangle_3+|-\rangle_1|+\rangle_2|+\rangle_3$ (one can easily show that $G^{(2)}(\theta_1,\theta_2)\propto 2+2\cos\{k\Delta(\cos\theta_1-\cos\theta_2)\}$). Fig. \ref{fig5b} shows  examples of post-semiclassical patterns of $G^{(2)}(\theta_1,\theta_2)$  obtained with the same initial state $|\psi\rangle$
at a specific time (see Appendix). Patterns are averaged over 100 realizations with correlated (a) and uncorrelated (b) noise patterns.  As shown in Fig. \ref{fig5a} and Fig. \ref{fig5b}(a), for the ``mirrored" realization of  polarization noise for different initial states, $G^{(2)}(\theta_1,\theta_2)$ patterns for quantum and post-semiclassical cases are identical.  The noise is "mirrored" when the $j$th TLS in the antenna array for the first initial state and the $N-j$th TLS of the array for the second initial state sense the same noise; see also Fig. \ref{pulses}(b) in the Appendix.   As shown in Fig. \ref{fig5b}(b),  uncorrelated noise sources lead to a semiclassical $G^{(2)}(\theta_1,\theta_2)$ pattern with a conserved position of the maxima compared to the full quantum case, but with distortions inducing a symmetry breaking.

So, we have demonstrated that it is indeed possible to use semiclassical antenna models for designing the higher-order correlation functions of the emitted field. The semiclassical approach can potentially serve as a handy modeling tool, since the number of equations to solve scales linearly with the number of the TLSs.

\section{Conclusions}

We have shown that it is possible to shape the second and higher order  correlation functions of the field emitted by a quantum antenna in the far-field zone by designing its initial state. We have proposed an optimization method using constrained linear and non-linear programming. We have demonstrated the feasibility of the method for designing states with two initial excitations. We have found states leading to highly co- or contra-directional emission of photon pairs for the same antenna, or even producing the effect of non-radiating sources by suppressing the field in the far-field zone. We have also shown that a quantum antenna can produce multi-photons momentum-entangled states. Despite the general quantum character of the state expected to produce desired spatial patterns of the correlation functions, we have also demonstrated that one can still use an appropriately modified semiclassical approach for this purpose. We believe that our method for producing patterned higher-order correlation functions of the emitted field can be of importance for imaging and high-precision sensing, as well as for designing an emitter-field interface for quantum information processing \cite{boto,rozema,supertwin,feber,chec}.

A. M., D. M., I. K.,  G. B. and D. B. acknowledge support from the EU project Horizon-2020 SUPERTWIN id.686731,
the National Academy of Sciences of Belarus program "Convergence", and the BRRFI project F17Y-004. G. Ya. S. acknowledges support from the project FP7-612285 CANTOR. G. B. and D. B. acknowledge Philippe Renevey for advices in coding.

\appendix

\begin{figure}
\begin{center}
\includegraphics[width=\linewidth]{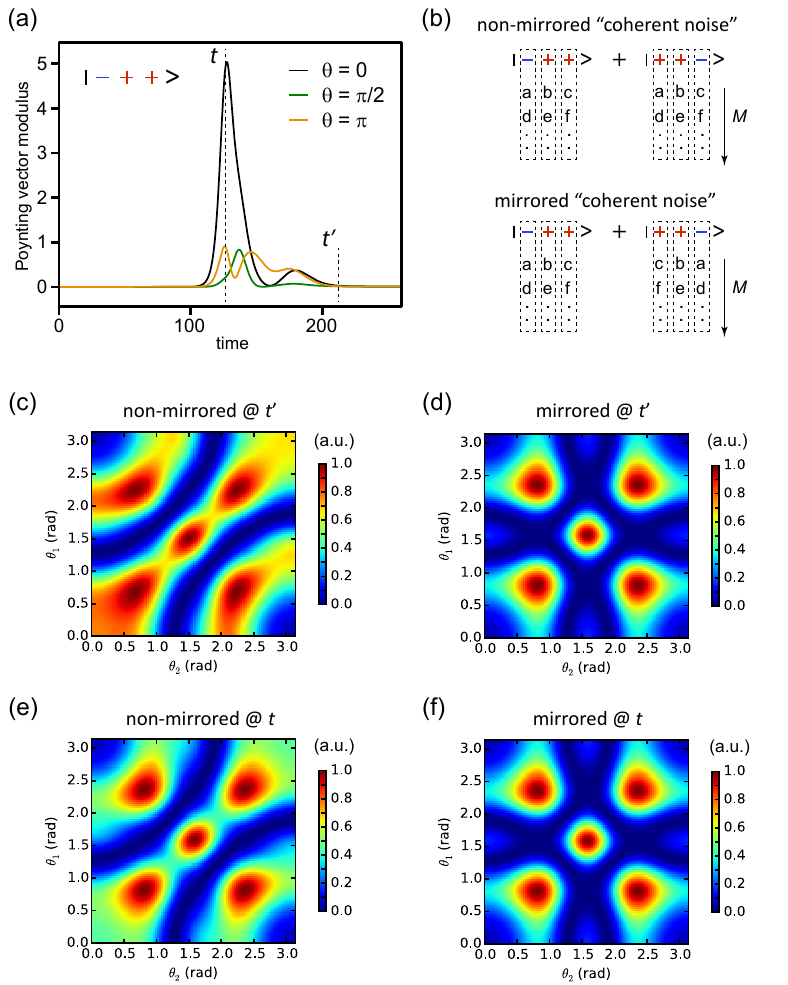}
\caption{\label{pulses} (a) Example of a superradiant pulse emitted by a linear chain of 3 dipoles with the initial conditions $\mathcal{Z}^{(1)}(0)=-1$, $\mathcal{Z}^{(2)}(0)=\mathcal{Z}^{(3)}(0)=1$ and $\mathcal{R}^{(1)}(0)=\mathcal{R}^{(2)}(0)=\mathcal{R}^{(3)}(0)=0$ for $\kappa\Delta=4.5$. Three observation angles $\theta=\{0, \pi/2, \pi \}$ are considered. (b) Schematics of the noise sources configurations for the non-mirrored and mirrored cases. Letters a, b, c, \textit{etc.} correspond to different noise patterns . (c)-(f) Different normalized $G^{(2)}(\theta_{1},\theta_{2})$ patterns for both noise configurations and at times $t$ and $t'$ in panel (a), each for $100$ realizations.}
\end{center}
\end{figure}

\section*{Appendix: semiclassical Maxwell-Bloch model of interacting dipoles}

We consider a regular linear chain of atoms modelled with a system of $N$ two-level atoms at equidistant positions in space given by the position vectors $\vec{R}_{j}$ (see Fig. \ref{fig1} of the main text). A semiclassical treatment of such a system, making use of the SVEA and RWA approximations, leads to a set of so-called Maxwell-Bloch equation (see $e.g.$ chapters 1 and 6 in Ref.~\cite{benedict}). Introducing relaxation processes and spontaneous fluctuations of the electric polarization in the medium under the form of a Langevin force term at each dipole (noise sources), we obtain the set of  differential equations below:

\begin{eqnarray}
\nonumber
\frac{d}{d\tau}\mathcal{R}^{(j)}(\tau) = -i\frac{\vec{d}_{j}}{\hbar}\vec{E}_{j}(\tau)\mathcal{Z}^{(j)}(\tau)-\\
\nonumber
\frac{\mathcal{R}^{(j)}(\tau)}{\tau_{2}}+\mathcal{L}^{j}(\tau), \\
\nonumber
\frac{d}{d\tau}\mathcal{Z}^{(j)}(\tau) = \frac{1}{2}\left(i\frac{\vec{d}_{j}}{\hbar}\vec{E}_{j}(\tau)\mathcal{R}^{(j)*}(\tau) + c.c.\right)-\\
\frac{1+\mathcal{Z}^{(j)}(\tau)}{\tau_{1}}
\label{MB}
\end{eqnarray}
where $\mathcal{Z}^{(j)}(\tau)$ and $\mathcal{R}^{(j)}(\tau)$ correspond to the population difference and polarization (or coherence) of the $j$th atom, respectively, and also  to the diagonal and off-diagonal elements of the single-atom density matrix, respectively.
$\mathcal{L}^{(j)}(\tau)$ is the Langevin force term (noise source) applied to the $j$th atom. $\tau_{1}$ and $\tau_{2}$ are the  relaxation times for inversion and macroscopic polarization, respectively. $\vec{d}_{j}$ is the transition dipole moment of the $j$th atom, and $\vec{{E}}_{j}(\tau)$ is the  electric field acting on the $j$th atom at the position $\vec{R}_{j}$. This field is a superposition of the microscopic fields $\vec{{E}}_{lj}(\tau)$ produced at the point $\vec{R}_{j}$ by all other  atoms labeled with index $l$, reading:

\begin{equation}
\vec{E}_{j}(\tau)=\sum_{l\neq j}\vec{E}_{lj}(\tau)
\end{equation}
where the amplitudes $\vec{{E}}_{lj}(\tau)$ are given by:

\begin{align}
\begin{split}
\vec{E}_{lj}(\tau) = \left[ \frac{3}{\Delta_{lj}^{3}} -\frac{3ik}{\Delta_{lj}^{2}} - \frac{k^{2}}{\Delta_{lj}} \right] \times \\
 (\vec{d}_{j} \vec{n}_{lj}) \vec{n}_{lj}\mathcal{R}^{(l)}(\tau) e^{ik\Delta_{lj}}\\
 - \left[ \frac{1}{\Delta_{lj}^{3}} -\frac{ik}{\Delta_{lj}^{2}} - \frac{k^{2}}{\Delta_{lj}} \right]
 {\vec{d}_{j}}  \mathcal{R}^{(l)}(\tau) e^{ik\Delta_{lj}}.
\end{split}\label{field}
\end{align}
Note that here we neglect the retardation in the amplitudes $\mathcal{R}^{(j)}(\tau)$, since we assume that the time for light to propagate through the system, $L/c$, is shorter than the characteristic superradiance (SR) time $T_{R}$, which defines the instability increment and the  the growth rate of the collective superradiant pulse \cite{boiko}. To favour  SR emission from our system of interacting TLS,  the relaxation time $\tau_1$ and decoherence time $\tau_{2}$ used in the modelling are much longer than $T_R$.  As one more   simplification, we consider dipole matrix elements $\vec{d}_\mathrm{l}$ of individual  TLS pointing up normally with respect to the chain axis (as it is assumed in the main text), so the first term on the right-hand side of expression~(\ref{field}) vanishes.

The component of the classical Poynting vector along the antenna axis takes the form:
\begin{equation}
S(\boldsymbol{r},\tau) \propto \sum_{l,m} \mathcal{R}^{(l)*}(\tau)\mathcal{R}^{(m)}(\tau) e^{ik\Delta_{lm}cos(\theta)}
\end{equation}
where $\theta$ is the angle between the linear array antenna and the direction of observation, and $\tau$ is the time a the observer position.

An example of a superradiant pulse emitted by a linear chain of 3 dipoles with the initial conditions $\mathcal{Z}^{(1)}(0)=-1$, $\mathcal{Z}^{(2)}(0)=\mathcal{Z}^{(3)}(0)=1$ and $\mathcal{R}^{(1)}(0)=\mathcal{R}^{(2)}(0)=\mathcal{R}^{(3)}(0)=0$ is shown in Fig.~\ref{pulses} for $\kappa\Delta=4.5$,  and $\mathcal{L}^{(k)}(\tau)$ being a zero-mean Gaussian noise. Three observation angles $\theta=\{0, \pi/2, \pi \}$ are considered.

It can be shown that the second-order correlation function $G^{2}(\theta_{1},\theta_{2})$ for the semiclassical case given by Eq. (\ref{gp}) in the main text takes here the form:
$G^{2}(\theta_{1},\theta_{2}) = \left| \mathcal{A}(\theta_{1},\theta_{2})\right|^{2}$, where the amplitude $\mathcal{A}(\theta_{1},\theta_{2})$ is given as:
\begin{eqnarray}
\nonumber
\mathcal{A}(\theta_{1},\theta_{2})=\left[ \mathcal{R}^{(1)} \mathcal{R}^{(2)}  \right]_{++-} \left( e^{ik\Delta \cos(\theta_{1})} + e^{ik\Delta \cos(\theta_{2})} \right) + \\
\nonumber
\left[ \mathcal{R}^{(2)} \mathcal{R}^{(3)}\right]_{-++} \times \\
\nonumber
\left( e^{ik\Delta (2 \cos(\theta_{1}) + \cos(\theta_{2}))} + e^{i k\Delta (\cos(\theta_{1}) + 2 \cos(\theta_{2}))} \right),
\end{eqnarray}
where $[\mathcal{R}^{(j)}]_{++-}$ and $[\mathcal{R}^{(j)}]_{-++}$ stand for the values of  the electric polarization of the $j$th atom for the initial states $|+\rangle_1|+\rangle_2|-\rangle_3$ and
$|-\rangle_1|+\rangle_2|+\rangle_3$, respectively. Two different configurations of the noise source terms $\mathcal{L}^{(j)}(\tau)$ are considered with the so-called non-mirrored and mirrored coherent noise, as shown in Fig.~\ref{pulses}(b).

Examples of normalized $G^{2}(\theta_{1},\theta_{2})$ functions calculated at two different times $t$ and $t'$ for both noise configurations are displayed in panels (c)-(f) for an averaging over $100$ realizations. It can be shown that for the mirrored noise case, the pattern is time independent and is identical to the pure quantum case shown in Fig.(6) in the main text. For the non-mirrored case, the patterns are time dependent but their variation over time is small and the position of the maxima is conserved compared to the pure quantum case.


\end{document}